\documentclass[a4paper,UKenglish]{lipicsnew}

\usepackage{epic,eepic,color,latexsym,amsmath,amssymb}

\newcommand{\haut}[1]{\mbox{\raisebox{1ex}{\scriptsize \it #1}}}

\newcommand{\arrow}{\mbox{\small$\longrightarrow$}}
\newcommand{\arrowleft}{\mbox{\small$\longleftarrow$}}

\newcommand{\fleche}[1]{\mbox{$\mathop{\arrow}\limits^{#1}$}}
\newcommand{\nofleche}[1]{\mbox{$\ {\tiny/}\hspace*{-1.1em}\mathop{\arrow}\limits^{#1}$}}
\newcommand{\deriv}{\mbox{\small$\longleftrightarrow$}}
\newcommand{\inverse}[1]{\mbox{$\mathop{\arrowleft}\limits^{#1}$}}
\newcommand{\croix}{\mbox{\scriptsize $\times$}}

\newcommand{\inter}[1]{\hbox{{\rm [}\hskip -2pt {\rm [}{$#1$}{\rm ]}\hskip -2pt {\rm ]}}}
\newcommand{\interInd}[1]{\tiny\hbox{{\rm [}\hskip -3.15pt{\rm [}{$#1$}{\rm ]}
\hskip -3.5pt {\rm ]}}}
\newcommand{\interFootnote}[1]{\footnotesize\hbox{{\rm [}\hskip -1.5pt
{\rm [}{$#1$}{\rm ]}\hskip -1.5pt {\rm ]}}}
\newcommand{\relatif}{\mbox{$\mathbb{Z}$}}
\newcommand{\entier}{\mbox{$\mathbb{N}$}}

\newcommand{\InfSup}[1]{\mbox{$\mathop{<}\!{#1}\!\mathop{>}$}}
\newcommand{\InfSupInd}[1]{\tiny\mbox{$\mathop{<}\!{#1}\!\mathop{>}$}}

\title{On Cayley graphs of basic algebraic structures}
\titlerunning{Cayley graphs}
\author[1]{Didier Caucal}
\affil[1]{CNRS, LIGM, University Paris-East, France\\
  \texttt{didier.caucal@univ-mlv.fr}}
\authorrunning{D. Caucal}

\begin{document}

\maketitle

\begin{abstract}
{\noindent}We present simple graph-theoretic characterizations of Cayley graphs 
for monoids, semigroups and groups. 
We extend these characterizations to commutative monoids, semilattices, and 
abelian groups.
\end{abstract}

\section{Introduction}

\hspace*{1.5em}Arthur Cayley was the first to define in 1854 \cite{Ca0} the 
notion of a group as well as the table of its operation known as the Cayley 
table. 
To describe the structure of a group~\,$(G,\cdot)$, Cayley also introduced in 
1878 \cite{Ca1} the concept of graph for \,$G$ \,according to a generating 
subset \,$S$, namely the set of labeled oriented edges 
\,$g\ \fleche{s}\ g{\cdot}s$ \,for every \,$g$ \,of~\,$G$ \,and \,$s$ \,of 
\,$S$. 
Such a graph, called Cayley graph, is directed and labeled in \,$S$ \,(or an 
encoding of \,$S$ \,by symbols called letters or colors). 
A characterization of unlabeled and undirected Cayley graphs was given by 
Sabidussi in 1958 \cite{Sa}\,: an unlabeled and undirected graph is a 
Cayley graph if and only if we can find a group with a free and transitive 
action on the graph. 
Following a question asked by Hamkins in 2010 \cite{Ha}: `Which graphs are 
Cayley graphs?', we gave simple graph-theoretic characterizations of Cayley 
graphs for groups, as well as for left-cancellative and cancellative monoids 
\cite{Cau}. 
In this paper, we generalize this last characterization to Cayley graphs of
monoids, then to semigroups. 
We also strengthen all these characterizations to commutative monoids, 
semilattices and abelian groups.

To structurally characterize the Cayley graphs (of groups), we selected four 
basic properties of these graphs. First and by definition, any Cayley graph is 
deterministic: there are no two edges of the same source and label. 
The right-cancellative property of groups induces the co-determinism of their 
graphs: there are no two edges of the same target and label. 
The left-cancellative property of groups implies that their graphs are simple: 
there are no two edges of the same source and goal. 
Finally, any Cayley graph is according to a generating subset hence is 
connected: there is a chain from the identity element to any vertex. 
To these four basic conditions is added the well known symmetry property of 
vertex-transitivity: all the vertices are isomorphic. 
These five properties satisfied by the Cayley graphs are sufficient to 
characterize them \cite{Cau}. 
Similarly, we obtained a graph-theoretic characterization for the Cayley graphs 
of cancellative monoids: first, they are now rooted since there is a path 
from the identity element to any vertex, and then by relaxing the vertex 
transitivity  to the forward vertex-transitivity: all the vertices are 
accessible-isomorphic \,{\it i.e.} \,the induced subgraphs by vertex 
accessibility are isomorphic \cite{Cau}.

To characterize the Cayley graphs of all monoids (not necessarily 
cancellative), we must weaken the forward vertex-transitivity. 
We say that a vertex is propagating if there is a homomorphism from its 
accessible subgraph to the accessible subgraph from any vertex. 
Thus, the identity of a monoid is a propagating vertex for each of its 
Cayley graphs. The identity is also an out-simple vertex: it is not source 
of two edges with the same target.
Moreover, any Cayley graph is source-complete: for any label of the graph and 
from any vertex, there is at least one edge. 
These properties are sufficient to characterize the Cayley graphs of monoids: 
they are the deterministic and source-complete graphs with a propagating 
out-simple root. 
It follows a graph-theoretic characterization for the Cayley graphs of 
semigroups (see Theorem~\ref{Semigroup}) and of cancellative semigroups 
(see Theorem~\ref{CancelSemigroup}).

For the Cayley graphs of commutative monoids, we just have to add the condition 
that any vertex \,$s$ \,is locally commutative: for any path from \,$s$ 
\,labeled by (two letters) \,$ab$, there is a path from \,$s$ \,labeled by 
\,$ba$ \,of the same target. 
The locally-commutativity can be restricted to a single vertex: the Cayley 
graphs of commutative monoids are the deterministic and source-complete graphs 
with a locally commutative propagating out-simple root. 
It follows a graph-theoretic characterization for the Cayley graphs of 
semilattices (see Theorem~\ref{Semilattice}).
By extending to chains the vertex propagation, we can restrict the 
vertex-transitivity of a Cayley graph to the existence of a single propagating 
vertex: the Cayley graphs of (resp. abelian) groups are the deterministic and 
co-deterministic, simple and connected graphs with a chain-propagating 
(resp. and locally commutative) source and target-complete vertex.

\section{Directed labeled graphs}

{\indent}We recall some basic concepts on directed labeled graphs, especially 
the vertex-transitivity and the forward vertex-transitivity.\\[-0.5em]

Let \,$A$ \,be an arbitrary (finite or infinite) set.  We denote by \,$A^*$ 
\,the set of tuples (words) over \,$A$ \,(the free monoid generated by \,$A$) 
\,and by \,$\varepsilon$ \,the \,$0$-tuple (the identity element called the 
empty word). 
A directed \,$A$-{\it graph} \,$(V,G)$ \,is defined by a set \,$V$ \,of 
{\it vertices} \,and a subset \,$G \,\subseteq \,V{\croix}A{\croix}V$ \,of 
{\it edges}. 
Any edge \,$(s,a,t) \in G$ \,is from the {\it source} \,$s$ \,to the 
{\it target} \,$t$ \,with {\it label} \,$a$, and is also written by the 
{\it transition} \,$s\ \fleche{a}_G\ t$ \,or directly \,$s\ \fleche{a}\ t$ \,if 
\,$G$ \,is clear from the context. 
The sources and targets of edges form the set \,$V_G$ \,of 
{\it non-isolated vertices} \,of~\,$G$ \,and we denote by \,$A_G$ \,the set of 
edge labels:\\[0.25em]
\hspace*{2em}$V_G\ =\ \{\ s\ |\ \exists\ a,t\ \ (s\ \fleche{a}\ t \,\vee 
\,t\ \fleche{a}\ s)\ \}$ \ \ \ and \ \ \ 
$A_G\ =\ \{\ a\ |\ \exists\ s,t\ \ (s\ \fleche{a}\ t)\ \}$.\\[0.25em]
We say that \,$G$ \,is {\it finitely labeled} \,if \,$A_G$ \,is finite. 
The set \,$V - V_G$ \,is the set of {\it isolated vertices}. 
From now on, we assume that any graph \,$(V,G)$ \,is without isolated vertex 
\,({\it i.e.} $V = V_G$), hence the graph can be identified with its edge 
set~\,$G$. We also exclude the empty graph \,$\emptyset$\,: every graph is a 
non-empty set of labeled edges. 
A vertex \,$s$ \,is an {\it out-simple vertex} \,if there are no two 
edges of source \,$s$ \,with the same target: 
\,$(s\ \fleche{a}\ t \,\wedge \,s\ \fleche{b}\ t)\ \Longrightarrow\ a=b$. 
A graph is {\it simple} \,if all its vertices are out-simple. 
We also say that \,$s$ \,is an {\it in-simple vertex} \,if there are no two 
edges with the same source and target \,$s$\,: 
\,$(t\ \fleche{a}\ s \,\wedge \,t\ \fleche{b}\ s)\ \Longrightarrow\ a=b$. 
Thus an in-simple vertex for \,$G$ \,is an out-simple vertex for 
\,$G^{-1}\ =\ \{\ (t,a,s)\ |\ (s,a,t) \in G\ \}$ \,the {\it inverse} of \,$G$. 
The {\it vertex-restriction} \,$G_{|P}$ \,of \,$G$ \,to a set \,$P$ \,is the 
induced subgraph of~\,$G$ \,by \,$P \cap V_G$\,:\\[0.25em]
\hspace*{12em}$G_{|P}\ =\ \{\ (s,a,t) \in G\ |\ s,t \in P\ \}$.\\[0.25em]
The {\it label-restriction} \,$G^{|P}$ \,of \,$G$ \,to a set \,$P$ \,is the 
subset of all its edges labeled in \,$P$\,:\\[0.25em]
\hspace*{12em}$G^{|P}\ =\ \{\ (s,a,t) \in G\ |\ a \in P\ \}$.\\[0.25em]
Let \,$\fleche{}_G$ \,be the unlabeled edge relation \,{\it i.e.}
\,$s\ \fleche{}_G\ t$ \,if \,$s\ \fleche{a}_G\ t$ \,for some \,$a \in A$. 
We write \,$\deriv_G$ \,for the unlabeled {\it adjacency relation} 
\,$\fleche{}_{G \cup G^{-1}}$ \,{\it i.e.} \,$s\ \deriv_G\ t$ \,for 
\,$s\ \fleche{}_G\ t$ \,or \,$t\ \fleche{}_G\ s$. 
We denote by \,$\fleche{}_G(s) \,= \,\{\ t\ |\ s\ \fleche{}_G\ t\ \}$ \,the set 
of {\it successors} \,of \,$s \in V_G$\,. 
We write \,$s\ \nofleche{}{}\,\!_G\ t$ \,if there is no edge in \,$G$ \,from 
\,$s$ \,to \,$t$ \,{\it i.e.} 
\,$G \,\cap \,\{s\}{\croix}A{\croix}\{t\} \,= \,\emptyset$. 
The {\it accessibility} \,relation 
\,$\fleche{}^*_G \,= \,\bigcup_{n \geq 0}\fleche{}^n_G$ \,is the reflexive and 
transitive closure under composition of \,$\fleche{}_G$\,. 
A graph \,$G$ \,is {\it accessible} \,from \,$P \subseteq V_G$ \,if for any
\,$s \in V_G$\,, there is \,$r \in P$ \,such that \,$r\ \fleche{}^*_G\ s$. 
We denote by \,$G_{{\downarrow}P}$ \,the induced subgraph of \,$G$ \,to the 
vertices accessible from \,$P$ \,which is the greatest subgraph of \,$G$ 
\,accessible from \,$P$. 
A {\it root} \,$r$ \,is a vertex from which \,$G$ \,is accessible \,{\it i.e.} 
\,$G_{{\downarrow}\{r\}}$ \,also denoted by \,$G_{{\downarrow}r}$ \,is equal to 
\,$G$. A {\it co-root} \,of \,$G$ \,is a root of \,$G^{-1}$. 
A graph \,$G$ \,is {\it strongly connected} \,if every vertex is a root: 
\,$s\ \fleche{}^*_G\ t$ \ \ for all~\,$s,t \in V_G$\,. 
A vertex \,$r$ \,of a graph \,$G$ \,is an \,$1${\it -root} \,if 
\,$r\ \fleche{}_G\ s$ \,for any vertex \,$s$ \,of \,$G$. 
A graph is {\it complete} if all its vertices are $1$-roots \,{\it i.e.} 
\,there is an edge between any couple of vertices: 
$\forall\ s,t \in V_G\ \ \exists\ a \in A_G \ (s\ \fleche{a}_G\ t)$. 
An $1${\it -coroot} \,of \,$G$ \,is an $1$-root of \,$G^{-1}$.\\
A graph \,$G$ \,is {\it co-accessible} \,from \,$P \subseteq V_G$ \,if 
\,$G^{-1}$ \,is accessible from \,$P$. 
A graph \,$G$ \,is {\it connected} \,if \,$G \,\cup \,G^{-1}$ \,is strongly 
connected.\\ 
A {\it path} \,$(s_0,a_1,s_1,\ldots,a_n,s_n)$ \,of {\it length} \,$n \geq 0$
\,in a graph \,$G$ \,is a sequence 
\,$s_0\ \fleche{a_1}\ s_1 \ldots \fleche{a_n}\ s_n$ \,of \,$n$ \,consecutive 
edges, and we write \,$s_0\ \fleche{a_1{\ldots}a_n}\ s_n$ \,for indicating the 
source \,$s_0$\,, \,the target~\,$s_n$ \,and the label word 
\,$a_1{\ldots}a_n \in A_G^*$ \,of the path; such a path is {\it elementary} 
\,if it goes through distinct vertices: \,$s_0 \neq \ldots \neq s_n$ \,and we 
write \,$s_0\ _{\neq}\!\!\fleche{a_1{\ldots}a_n}\ s_n$\,. 
We write \,$s\ \fleche{u,v}_G\ t$ \,if \,$s\ \fleche{u}_G\ t$ \,and 
\,$s\ \fleche{v}_G\ t$\,; \,we also denote by \,$s\ \fleche{u,v}_G$ \,if 
\,$s\ \fleche{u,v}_G\ t$ \,for some \,$t$.\\
Let \,$u = a_1{\ldots}a_m$ \,and \,$v = b_1{\ldots}b_n$ \,where \,$m, n \geq 0$ 
\,and \,$a_1,\ldots,a_m,b_1,\ldots,b_n \in A$. 
We write \,$s\ _{\neq}\fleche{u,v}_G$ \,if \,$uv \neq \varepsilon$ \,and there 
exists paths 
\,$s\ \fleche{a_1}_G\ s_1 \ldots \fleche{a_m}\ s_m$ 
\,and~\,$s\ \fleche{b_1}_G\ t_1 \ldots \fleche{b_n}\ t_n~=~s_m$ \,forming an 
elementary cycle: 
\,$s \neq s_ 1\neq \ldots \neq s_m \neq t_1 \neq \ldots \neq t_{n-1}$ \,which 
is  illustrated as follows:
\begin{center}
\includegraphics{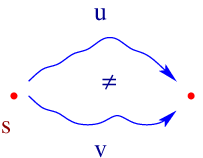}\\
\end{center}
In particular for a loop \,$s\ \fleche{a}\ s$, we have 
\,$s\ _{\neq}\fleche{\varepsilon,a}$ \,and for two edges 
\,$s\ \fleche{a}\ t$ \,and \,$s\ \fleche{b}\ t$ \,of the same source and goal, 
we have \,$s\ _{\neq}\fleche{a,b}$\,.\\
Recall that a {\it morphism} \,from a graph \,$G$ \,into a graph \,$H$ \,is a 
mapping \,$h$ \,from \,$V_G$ \,into~\,$V_H$ \,such that 
\,$s\ \fleche{a}_G\ t \ \Longrightarrow \ h(s)\ \fleche{a}_H\ h(t)$; we write 
\,$G\ \fleche{}_h\ H$. 
If, in addition, $h$ \,is bijective and \,$h^{-1}$ \,is a morphism, $h$ \,is 
called an {\it isomorphism} \,from \,$G$ \,to \,$H$\,; \,we write 
\,$G \,\equiv_h \,H$ \,or directly \,$G \,\equiv \,H$ \,if we do not specify 
an isomorphism, and we say that \,$G$ \,and \,$H$ \,are {\it isomorphic}. 
An {\it automorphism} \,of \,$G$ \,is an isomorphism from \,$G$ \,to \,$G$.
\begin{lemma}\label{Isomorphism}
For any deterministic graphs \,$G$ \,and \,$H$ \,rooted respectively by \,$s$ 
\,and \,$t$,\\
\hspace*{6em}if \ $G\ \fleche{}_g\ H$ \,and \,$H\ \fleche{}_h\ G$ \,with 
\,$g(s) = t$ \,and \,$h(t) = s$ \ then \ $G \,\equiv_g \,H$.
\end{lemma}
\proof\mbox{}\\
{\bf i)} Let us check that \,$h(g(x)) = x$ \,for any \,$x \in V_G$\,.\\
The proof is done by induction on 
\,$\ell(x) \,= \,{\rm min}\{\ n\ |\ s\ \fleche{}^n_G\ x\ \}$.\\
$\ell(x) = 0$\,: \,$x = s$ \,and \,$h(g(s)) \,= \,h(t) \,= \,s$.\\
$\ell(x) > 0$\,: \,There is \,$x'\ \fleche{a}\ x$ \,such that 
\,$\ell(x') = \ell(x)-1$. 
As \,$g$ \,is a morphism, $g(x')\ \fleche{a}\ g(x)$.\\
By induction hypothesis and as \,$h$ \,is a morphism, 
$x' \,= \,h(g(x'))\ \fleche{a}\ h(g(x))$.\\
As \,$G$ \,is deterministic, we get \,$h(g(x)) = x$.\\[0.25em]
{\bf ii)} By (i), $g$ \,is injective. 
By exchanging \,$g$ \,with \,$h$ and by (i), 
$g(h(y)) = y$ \,for any \,$y \in V_H$\,.\\
In particular \,$g$ \,is surjective. Thus \,$g$ \,is bijective and 
\,$h = g^{-1}$. So \,$G \,\equiv_g \,H$.
\qed\\[1em]
In Lemma~\ref{Isomorphism}, even if \,$g$ \,and \,$h$ \,are surjective, 
the condition \,$g(s) = t$ \,and \,$h(t) = s$ \,is necessary. 
For instance, the two non-isomorphic graphs below are rooted, deterministic, 
co-deterministic, and there is a surjective morphism from one into the other.
\begin{center}
\includegraphics{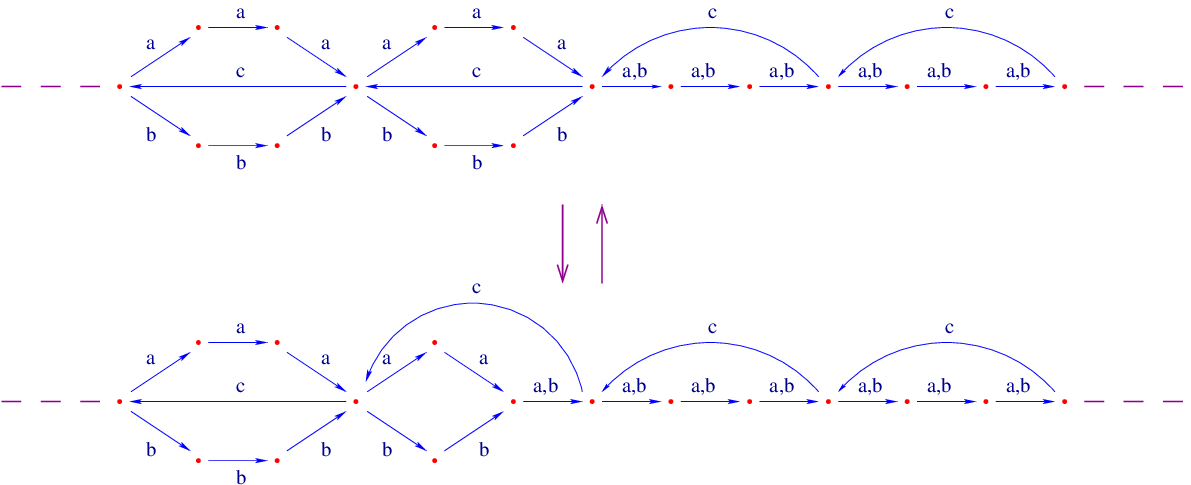}\\
\end{center}
Two vertices \,$s,t$ \,of a graph \,$G$ \,are {\it isomorphic} 
\,and we write \,$s \,\simeq_G t$ \,if \,$t = h(s)$ \,for some automorphism 
\,$h$ \,of \,$G$. A graph \,$G$ \,is {\it vertex-transitive} \,if all the 
vertices are isomorphic \,{\it i.e.} \,$s \,\simeq_G t$ \,for every 
\,$s,t \in V_G$\,. 
Two vertices \,$s,t$ \,of a graph \,$G$ \,are {\it accessible-isomorphic} 
\,and we write \,$s \downarrow_G t$ \,if \,$t = h(s)$ \,for some isomorphism 
\,$h$ \,from \,$G_{{\downarrow}s}$ \,to \,$G_{{\downarrow}t}$\,. 
A graph \,$G$ \,is {\it forward vertex-transitive} \,if all its vertices are 
accessible-isomorphic: \,$s \downarrow_G t$ \,for every \,$s,t \in V_G$\,. 
Any vertex-transitive graph is forward vertex-transitive. 
The semiline \,$\{\ n\ \fleche{a}\ n+1\ |\ n \in \entier\ \}$ \,is 
forward vertex-transitive but is not vertex-transitive. 
Any strongly connected forward vertex-transitive graph is vertex-transitive.\\
We need to circulate in a graph \,$G$ \,in the direct and inverse direction of 
the arrows.\\
Let \,$\haut{\mbox{---}} : A_G\ \fleche{}\ A - A_G$ \,be an injective mapping 
of image \,$\overline{A_G} \,= \,\{\ \overline{a}\ |\ a \in A_G\ \}$ \,a 
disjoint copy of \,$A_G$\,. This allows to define the graph\\[0.25em]
\hspace*{10em}$\overline{G} \ = \ G \,\cup \,\{\ t\ \fleche{\overline{a}}\ s\ |
\ s\ \fleche{a}_G\ t\ \}$\\[0.25em]
in order that \,$V_{\overline{G}} = V_G$ \,and \,$A_{\overline{G}} \,= \,A_G \cup 
\overline{A_G}$ \,with\\[0.25em]
\hspace*{1.5em}\begin{tabular}{rcl}
$G$ \,deterministic and co-deterministic & $\Longrightarrow$ & $\overline{G}$ 
\,deterministic and co-deterministic\\
$G$ \,source and target-complete & $\Longrightarrow$ & $\overline{G}$ 
\,source and target-complete\\  
$G$ \,connected & $\Longrightarrow$ & $\overline{G}$ \,strongly connected\\
$s \,\simeq_G t$ & $\Longleftrightarrow$ & $s \,\downarrow_{\overline{G}} \,t$ \ 
for any \,$s,t \in V_G$
\end{tabular}\\[0.25em]
hence \,$G$ \,is vertex-transitive if and only if \,$\overline{G}$ \,is forward 
vertex-transitive.\\
A path of \,$\overline{G}$ \,{\it i.e.} \,$s\ \fleche{u}_{\overline{G}}\ t$ 
\,with \,$u \in (A_G \cup \overline{A_G})^*$ \,is a {\it chain} \,of \,$G$ 
\,also denoted by \,$s\ \fleche{u}_G\ t$ \,where 
\,$s\ \fleche{\overline{a}}_G\ t$ \,means that \,$t\ \fleche{a}_G\ s$ \,for any 
\,$a \in A_G$\,. Thus\\[0.25em]
\hspace*{10em}$s\ \fleche{u}_G\ t \ \ \ \Longleftrightarrow \ \ \ 
t\ \fleche{\widetilde{\overline{u}}}_G\ s$ \ \ for any 
\,$u \in (A_G \cup \overline{A_G})^*$\\[0.25em]
such that for \,$u = a_1{\ldots}a_n$ \,with \,$n \geq 0$ \,and 
\,$a_1,\ldots,a_n \in A_G \cup \overline{A_G}$\,, 
\,$\overline{u} \,= \,\overline{a_1}\ldots\overline{a_n}$ \,where 
\,$\overline{\overline{a}} = a$ \,for any \,$a \in A_G \cup \overline{A_G}$\,, 
\,and \,$\widetilde{u} \,= \,a_n{\ldots}a_1$~\,is the {\it mirror} \,of \,$u$.
\\ 
Let us give basic properties on (forward) vertex-transitive graphs.
\begin{fact}\label{SourceComplet}
Any forward vertex-transitive graph is source-complete.\\
\hspace*{4.9em}Any vertex-transitive graph is source and target-complete.  
\end{fact}
The forward vertex-transitivity of a rooted graph is reduced to the 
accessible-isomorphism of a root with its successors. 
The vertex-transitivity of a connected graph is reduced to the isomorphism of 
a vertex with its adjacent vertices.
\begin{lemma}\label{AccessTransitive}
A graph \,$G$ \,of root \,$r$ \,is forward vertex-transitive \,iff 
\ $r \downarrow_G s$ \,for any \,$r\ \fleche{}_G\ s$.\\
\hspace*{6em}A connected graph with a vertex \,$r$ \,is vertex-transitive 
\,{\rm if and only if}\\
\hspace*{12em}$r \,\simeq_G \,s$ \,for any \,$r\ \deriv{}_{G}\ s$.
\end{lemma}
\proof\mbox{}\\
Let \,$G$ \,be a graph with a root \,$r$ \,such that
\,$r \downarrow_G s$ \,for any \,$r\ \fleche{}_G\ s$.\\
Let us check that \,$G$ \,is forward vertex-transitive \,{\it i.e.}
\,$r \downarrow_G s$ \,for any \,$r\ \fleche{}_G^*\ s$.\\
The proof is done by induction on \,$n \geq 0$ \,for \,$r\ \fleche{}_G^n\ s$.\\
For \,$n = 0$, we have \,$r = s$. For \,$n > 0$, let \,$t$ \,be a vertex such 
that \,$r\ \fleche{}_G^{n-1}\ t\ \fleche{}_G\ s$.\\
By induction hypothesis, we have \,$r \downarrow_G t$ \,{\it i.e.} 
\,$f(r) = t$ \,for some isomorphism \,$f$ \,from \,$G_{{\downarrow}r}$ \,to 
\,$G_{{\downarrow}t}$\,. 
As \,$t\ \fleche{}_G\ s$, there exists \,$r'$ \,such that 
\,$r\ \fleche{}_G\ r'$ \,and \,$f(r') = s$. So \,$r' \downarrow_G s$.\\
By hypothesis \,$r \downarrow_G r'$. 
By transitivity of \,$\downarrow_G$\,, we get \,$r \downarrow_G s$.\\
We get the second equivalence using the first one for \,$\overline{G}$.
\qed

\section{Commutative and propagating graphs}

{\indent}We recall when a graph is deterministic, co-deterministic, 
source-complete, target-complete, commutative. 
All these notions are equivalent when they are defined globally by paths or 
locally by edges. 
We introduce the propagation of joined paths which allows to express 
differently accessible-isomorphic vertices for deterministic graphs. 
The propagation can be restricted to elementary paths for deterministic and 
source-complete graphs. 
Finally we extend to chains the commutation and the propagation.\\[-0.5em]

A graph is {\it deterministic} \,if there are no two paths with the same source 
and label word:\\
\hspace*{6em}$(r\ \fleche{u}\ s \,\wedge \,r\ \fleche{u}\ t)\ \Longrightarrow\ 
s=t$ \ for any \,$r,s,t \in V_G$ \,and \,$u \in A_G^*$\,.\\
This definition coincides with the local property that there are no two edges 
with the same source and label: 
\,$(r\ \fleche{a}\ s \,\wedge \,r\ \fleche{a}\ t)\ \Longrightarrow\ 
s=t$ \ for any \,$r,s,t \in V_G$ \,and \,$a \in A_G$\,.\\
A graph is {\it co-deterministic} \,if its inverse is deterministic:
there are no two paths (resp. edges) with the same target and label word 
(resp. label).\\
A graph \,$G$ \,is a {\it source-complete graph} \,if for all vertex \,$s$
\,and label word \,$u$, there exists a path from \,$s$ \,labeled by 
\,$u$\,: \,$\forall\ s \in V_G\ \ \forall\ u \in A_G^*\ \ \exists\ t \in V_G \ 
(s\ \fleche{u}_G\ t)$. 
Locally a vertex \,$r$ \,is a {\it source-complete vertex} \,if for any label 
\,$a \in A_G$ \,there exists \,$s \in V_G$ \,such that \,$r\ \fleche{a}_G\ s$. 
Thus\\
\hspace*{6em}$G$ \,is source-complete \ \ $\Longleftrightarrow$ \ \ 
all its vertices are source-complete.\\
Similarly a vertex \,$r$ \,of a graph \,$G$ \,is a 
{\it target-complete vertex} \,if \,$r$ \,is source-complete for \,$G^{-1}$ 
\,{\it i.e.} \,$\forall\ a \in A_G\ \ \exists\ s \in V_G \ 
(s\ \fleche{a}_G\ r)$. 
A graph \,$G$ \,is a {\it target-complete graph} \,if its inverse is 
source-complete \,{\it i.e.} \,all the vertices of \,$G$ \,are target-complete.
\\
Let us recall the path commutation in a graph. 
Let \,$\approx_{A}$ \,be the binary {\it commutative relation} \,on \,$A^*$ 
\,defined by \,$uabv \approx_A ubav$ \,for any \,$a,b \in A$ \,and 
\,$u,v \in A^*$. By reflexivity and transitivity, we extend \,$\approx_A$ \,to 
the {\it commutation congruence} \,$\cong_A$\,. 
A vertex \,$r$ \,of a graph \,$G$ \,is a {\it commutative vertex} \,if 
\,$r\ \fleche{u}_G\ s \ \Longrightarrow \ r\ \fleche{v}_G\ s$ \,for any 
\,$s \in V_G$ \,and any \,$u,v \in A^*$ \,such that \,$u \,\cong_A \,v$\,. 
For the following deterministic and source-complete graph:
\begin{center}
\includegraphics{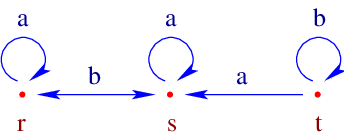}\\
\end{center}
the vertices \,$r$ \,and \,$s$ \,are commutative but \,$t$ \,is not a 
commutative since \,$t\ \fleche{ab}\ r$ \,and \,$t\ \fleche{ba}\ s$.\\
We say that \,$G$ \,is a {\it commutative graph} \,if all its vertices are 
commutative. 
For instance, the following deterministic graph is commutative:
\begin{center}
\includegraphics{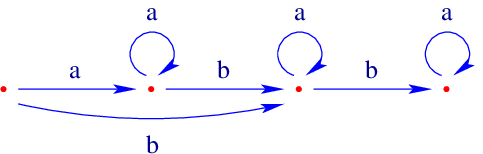}\\
\end{center}
Let us restrict this commutation from paths to edges. 
We say that a vertex \,$r$ \,of a graph~\,$G$ \,is a 
{\it locally commutative vertex} \,if 
\,$r\ \fleche{ab}_G\ s \ \Longrightarrow \ r\ \fleche{ba}_G\ s$ \,for any 
\,$s \in V_G$ \,and any \,$a,b \in A_G$\,.
The commutation of all vertices may be restricted to the local commutation.
\begin{lemma}\label{Commutative}
A graph is commutative {\rm if and only if} all its vertices are locally 
commutative.
\end{lemma}
\proof\mbox{}\\
$\Longrightarrow$\,: Any commutative vertex is locally commutative.\\
$\Longleftarrow$\,: Let a graph \,$G$ \,whose any vertex is locally 
commutative. 
Let a path \,$s\ \fleche{u}_G\ t$.\\
We have to check that \,$s\ \fleche{v}_G\ t$ \,for any \,$v \cong_A u$.\\
By induction on the minimum number of commutations between \,$u$ \,and \,$v$, 
we can restrict to \,$u \approx_A v$ \,{\it i.e.} \,$u = xaby$ \,and 
\,$v = xbay$ \,for some \,$a,b \in A_G$ \,and \,$x,y \in A_G^*$\,.\\
Thus \,$s\ \fleche{x}\ s'\ \fleche{ab}\ t'\ \fleche{y}\ t$ \,for some vertices 
\,$s',t'$.\\
As \,$s'$ \,is locally commutative, we have \,$s'\ \fleche{ba}\ t'$ \,hence 
\,$s\ \fleche{v}\ t$.
\qed\\[1em]
We will now express the accessible-isomorphism of vertices by propagation of 
confluent paths. We start with the propagation of loops. 
We say that a vertex \,$r$ \,of a graph \,$G$ \,is a 
{\it loop-propagating vertex} \,when we have the following property: 
if \,$r$ \,has a loop labeled by \,$a \in A$ \,then any vertex has a loop 
labeled by~\,$a$\,: \ 
$r\ \fleche{a}_G\ r \ \ \Longrightarrow \ \ \forall\ s \in V_G\ 
(s\ \fleche{a}_G\ s)$.
\begin{lemma}\label{LoopProp}
Any locally commutative $1$-root of a deterministic graph is loop-propagating.
\end{lemma}
\proof\mbox{}\\
Let \,$G$ \,be a deterministic graph and \,$r$ \,be a locally commutative 
\,$1$-root.\\
Let a loop \,$r\ \fleche{a}\ r$ \,and a vertex \,$s$. 
As \,$r$ \,is an \,$1$-root, there exists \,$b \in A$ \,such that 
\,$r\ \fleche{b}\ s$.\\
So \,$r\ \fleche{ab}\ s$. 
As \,$r$ \,is locally commutative, $r\ \fleche{ba}\ s$.
As \,$G$ \,is deterministic, we get \,$s\ \fleche{a}\ s$.
\qed\\[1em]
We extend the propagation of loops to paths.\\
A vertex \,$r$ \,is {\it propagating} \,(resp. $1${\it -propagating}) if for 
any \,$u,v \in A_G^*$ \,(resp. $A_G$),\\[0.25em]
\hspace*{10em}$r\ \fleche{u,v}_G \ \ \ \Longrightarrow \ \ \ 
\forall\ s \in V_G\ (s\ \fleche{u,v}_G\,)$.\\[0.25em]
The restriction of this implication to any \,$u \in A_G$ \,with 
\,$v = \varepsilon$ \,is the loop-propagating notion. 
The restriction of the implication to \,$u = v$ \,means that 
\,$r\ \fleche{u}_G \ \ \Longrightarrow \ \ \forall\ s \in V_G\ 
(s\ \fleche{u}_G\,)$.\\
In particular for the graph \,$\{r\ \fleche{a}\ s\}$, the vertex \,$r$ \,is 
not $1$-propagating and \,$s$ \,is propagating.\\
For the following deterministic and source-complete graph:
\begin{center}
\includegraphics{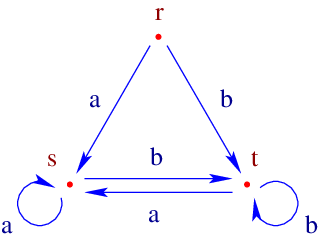}\\
\end{center}
the vertex \,$r$ \,is propagating but the vertices \,$s,t$ \,are not 
propagating: we have \,$s\ \fleche{\varepsilon,a}$ (resp. 
$t\ \fleche{\varepsilon,b}$) which is not the case for the other vertices. 
All the vertices are \,$1$-propagating.\\
For the following two deterministic connected graphs:
\begin{center}
\includegraphics{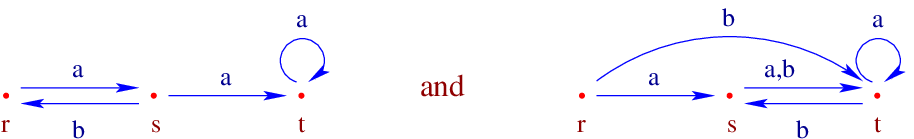}\\
\end{center}
the vertices \,$r,t$ \,are \,$1$-propagating (but not propagating) and \,$s$ 
\,is not \,$1$-propagating.\\
A propagating vertex of a deterministic graph is a vertex from which there is 
a morphism linking it to any vertex.
\begin{lemma}\label{PropagatingReducible}
For any deterministic graph \,$G$ \,and vertices \,$r,s$, we have\\
\hspace*{7.5em}$r\ \fleche{u,v}_G \ \Longrightarrow \ s\ \fleche{u,v}_G$ \ \ 
for any \,$u,v \in A_G^*$\\
\hspace*{1.2em}{\rm if and only if} \ there is a morphism \,$h$ \,from 
\,$G_{{\downarrow}r}$ \,to \,$G_{{\downarrow}s}$ \,such that \,$h(r) = s$.
\end{lemma}
\proof\mbox{}\\
$\Longleftarrow$\,: \,Immediate for any graph.\\
$\Longrightarrow$\,: \,As \,$G$ \,is deterministic, it allows to define the 
mapping \,$h : V_{G_{{\downarrow}r}}\ \fleche{}\ V_{G_{{\downarrow}s}}$ \,by\\
\hspace*{6em}$h(p) \,= \,q$ \ if \ $r\ \fleche{u}_G\ p$ \,and 
\,$s\ \fleche{u}_G\ q$ \,for some \,$u \in A^*$.\\
Thus \,$h(r) = s$. 
It remains to check that \,$h$ \,is a morphism.\\
Let \,$p\ \fleche{a}_{G_{{\downarrow}r}}\ q$. 
There exists \,$u \in A_G^*$ \,such that \,$r\ \fleche{u}_G\ p$.\\
As \,$r\ \fleche{ua}_G$\,, we have \,$s\ \fleche{ua}_G$ \,{\it i.e.} 
\,$s\ \fleche{u}_G\ p'\ \fleche{a}_G\ q'$ \,for some vertices \,$p',q'$.\\
As \,$G$ \,is deterministic, $h(p) = p'$ \,and \,$h(q) = q'$ \,hence 
\,$h(p)\ \fleche{a}_{G_{{\downarrow}s}}\ h(q)$.
\qed\\[1em]
The determinism condition in Lemma~\ref{PropagatingReducible} is necessary: 
for the following non deterministic (and non connected) graph \,$G$\,:
\begin{center}
\includegraphics{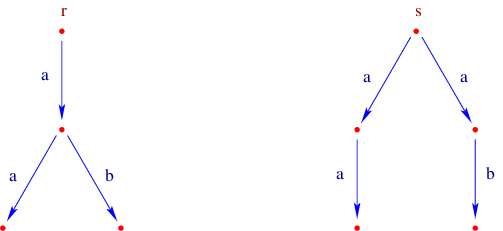}\\
\end{center}
we have \,$(r\ \fleche{u,v}_G \ \Longrightarrow \ s\ \fleche{u,v}_G)$ \,for any 
\,$u,v \in A_G^*$ \,and there is no morphism from \,$G_{{\downarrow}r}$ \,into 
\,$G$ \,linking \,$r$ \,to \,$s$. 
Let us give other basic properties on \,$1$-propagating vertices.
\begin{fact}\label{SourcePropagating}
Any graph with a source-complete $1$-propagating vertex is source-complete.\\
\hspace*{4.9em}For any source-complete graph, any out-simple vertex is 
$1$-propagating.
\end{fact}
Here is a source-complete and deterministic graph without $1$-propagating 
vertex.
\begin{center}
\includegraphics{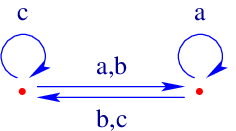}\\
\end{center}
The existence of a source-complete propagating vertex in a deterministic graph 
allows to reduce the commutativity of the graph to the locally commutativity 
of the vertex.
\begin{lemma}\label{CommutativeForward}
Let a deterministic graph \,$G$ \,with a source-complete and propagating 
vertex~$r$.\\
\hspace*{6em}If \,$r$ \,is locally commutative \,then \,$G$ \,is commutative.
\end{lemma}
\proof\mbox{}\\
We have to show that \,$G$ \,is commutative.\\
Let \,$s\ \fleche{ab}\ t$ \,with \,$a,b \in A$. 
By Lemma~\ref{Commutative}, it remains to check that \,$s\ \fleche{ba}\ t$.\\
By Fact~\ref{SourcePropagating}, $G$ \,is source-complete. 
So \,$r\ \fleche{ab}\ r'$ \,for some vertex \,$r'$.\\
As \,$r$ \,is locally commutative, $r\ \fleche{ba}\ r'$.\\
As \,$r\ \fleche{ab,ba}$ \,and \,$r$ \,is propagating, we get 
\,$s\ \fleche{ab,ba}\ t'$ \,for some vertex \,$t'$.\\
As \,$G$ \,is deterministic, $t = t'$ \,thus \,$s\ \fleche{ba}\ t$.
\qed\\[1em]
When a graph is deterministic, two vertices are accessible-isomorphic means 
that each vertex is propagating for the other.
\begin{lemma}\label{ForwardTransitive}
For any deterministic graph \,$G$ \,and any vertices \,$s,t$, we have\\
\hspace*{9em}$s \downarrow_G t \ \ \Longleftrightarrow \ \ 
\forall\ u,v \in A_G^*\ (s\ \fleche{u,v}_G \ \Longleftrightarrow \ 
t\ \fleche{u,v}_G\,)$.
\end{lemma}
\proof\mbox{}\\
$\Longrightarrow$\,: \,Obvious for any graph \,$G$.\\
$\Longleftarrow$\,: \ Let \,$s,t \in V_G$ \,such that \,$s\ \fleche{u,v}_G$ 
\,if and only if \,$t\ \fleche{u,v}_G$ \ for any \,$u,v \in A^*$.\\
As \,$G$ \,is deterministic, we define the mappings 
\,$g : V_{G_{{\downarrow}s}}\ \fleche{}\ V_{G_{{\downarrow}t}}$ \,and 
\,$h : V_{G_{{\downarrow}t}}\ \fleche{}\ V_{G_{{\downarrow}s}}$ \,by\\
\hspace*{3em}$g(p) \,= \,q$ \,and \,$h(q) = p$ \ if \ $s\ \fleche{u}_G\ p$ 
\,and \,$t\ \fleche{u}_G\ q$ \,for some \,$u \in A^*$.\\
As yet done in the proof of Lemma~\ref{PropagatingReducible}, $g$ \,and \,$h$ 
\,are morphisms with \,$g(s) = t$ \,and \,$h(t) = s$.\\
By Lemma~\ref{Isomorphism}, $g$ \,is an isomorphism from \,$G_{{\downarrow}s}$ 
\,to \,$G_{{\downarrow}t}$ \,{\it i.e.} \,$s \downarrow_G t$.
\qed\\[1em]
A graph \,$G$ \,is a {\it propagating graph} (resp. 
$1${\it -propagating graph}) if all its vertices are pro\-pagating (resp. 
$1$-propagating) \,{\it i.e.} \,for any \,$u,v \in A_G^*$ \,(resp. $A_G$) and 
any \,$s,t \in V_G$\,, \,$s\ \fleche{u,v}_G \ \ \Longrightarrow \ \ 
t\ \fleche{u,v}_G$\,. For the following two connected graphs:
\begin{center}
\includegraphics{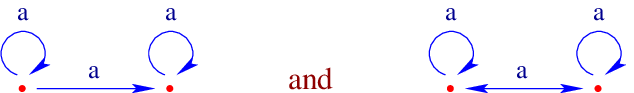}\\
\end{center}
the first graph is propagating but not forward vertex-transitive, and the 
second graph is forward vertex-transitive hence propagating. 
The $1$-propagating property and the source-complete property coincide for the 
simple graphs.
\begin{fact}\label{SourceCompletePropagating}
Any $1$-propagating graph is source-complete.\\
\hspace*{5.4em}Any $1$-propagating graph with an out-simple vertex is a simple 
graph.\\
\hspace*{5.4em}Any source-complete and simple graph is $1$-propagating.
\end{fact}
Let us restrict the path propagation to elementary paths.\\
A vertex \,$r$ \,is an {\it elementary-propagating vertex} \,if for any 
\,$u,v \in A_G^*$\,,\\[0.25em]
\hspace*{10em}$r\ _{\neq}\fleche{u,v}_G \ \ \ \Longrightarrow \ \ \ 
\forall\ s \in V_G\ (s\ _{\neq}\fleche{u,v}_G\,)$.\\[0.25em]
An {\it elementary-propagating graph} \,is a graph whose any vertex is 
elementary-propagating. 
For instance, the deterministic and simple graph \,$\{r\ \fleche{a}\ s\}$ \,is 
elementary-propagating but is not source-complete, hence is not 
$1$-propagating. 
The elementary-propagating property coincides with the propagating property 
for the deterministic and source-complete graphs.
\begin{proposition}\label{ForwardVertexTransitive}
{\bf a)} A source-complete graph \,$G$ \,is propagating \ {\rm if and only if}\\
\hspace*{9em}$s\ _{\neq}\fleche{u,v}_G \ \ \Longrightarrow \ \ 
t\ \fleche{u,v}_G\,$ \ for any \,$s,t \in V_G$ \,and \,$u,v \in A_G^*$\,.\\
\hspace*{1em}{\bf b)} A deterministic source-complete graph is propagating 
\,{\rm iff} \,it is elementary-propagating.\\
\hspace*{1em}{\bf c)} A deterministic graph is propagating 
\,{\rm if and only if} \,it is forward-vertex transitive.\\
\hspace*{1em}{\bf d)} A deterministic graph \,$G$ \,with a root \,$r$ \,is a 
propagating graph \ {\rm if and only if}\\
\hspace*{9.5em}$r\ \fleche{u,v}_G \ \ \Longleftrightarrow \ \ 
s\ \fleche{u,v}_G\,$ \ for any \,$r\ \fleche{}_G\ s$ \,and \,$u,v \in A_G^*$\,.
\\
\hspace*{1em}{\bf e)} A deterministic target-complete graph with a root \,$r$ 
\,is propagating \ {\rm iff} \ $r$ \,is propagating.
\end{proposition}
\proof\mbox{}\\
{\bf i)} Let us check (a).\\
$\Longrightarrow$\,: \,Immediate for any propagating graph.\\
$\Longleftarrow$\,: \,Assume that 
\,$s\ _{\neq}\fleche{u,v}_G \ \ \Longrightarrow \ \ 
t\ \fleche{u,v}_G\,$ \ for any \,$s,t \in V_G$ \,and \,$u,v \in A_G^*$\,.\\
Let us show that \,$s\ \fleche{u,v}_G \ \ \Longrightarrow \ \ 
t\ \fleche{u,v}_G\,$ \ for any \,$s,t \in V_G$ \,and \,$u,v \in A_G^*$\,.\\
The proof is done by induction on \,$|uv| \geq 0$.\\
$|uv| = 0$\,: \,$u = v = \varepsilon$ \,hence \,$t\ \fleche{u,v}_G\ t$ \,for 
any vertex \,$t$.\\
$|uv| > 0$\,: Let \,$s\ \fleche{u,v}_G$ \,with \,$u,v \in A_G^*$ \,and let 
\,$t \in V_G$\,. 
We can assume that \,$|u| \leq |v|$.\\
We distinguish the two complemantary cases below.\\
{\it Case 1}\,: \,$u = \varepsilon$. 
So \,$s\ _{\neq}\fleche{\varepsilon,v'}_G\ s\ \fleche{\varepsilon,v''}_G\,$ 
\,for some \,$v'v'' = v$.\\
\hspace*{1em}By hypothesis \,$t\ \fleche{\varepsilon,v'}_G\ t$. 
By induction hypothesis \,$t\ \fleche{\varepsilon,v''}_G\,$ 
\,hence \,$t\ \fleche{\varepsilon,v}_G$\,.\\
{\it Case 2}\,: \,$u \neq \varepsilon$. We distinguish the two subcases below.\\
\hspace*{1em}{\it Case 2.1}\,: \,$s\ \fleche{a}_G\ s'\ \fleche{u',v'}_G$ \,with 
\,$a \in A_G$\,, $u = au'$ \,and \,$v = av'$.\\
\hspace*{2em}As \,$G$ \,is source-complete, $t\ \fleche{a}_G\ t'$ \,for some 
vertex \,$t'$.\\
\hspace*{2em}By induction hypothesis, \,$t'\ \fleche{u',v'}_G$ \,hence 
\,$t\ \fleche{u,v}_G$\,.\\
\hspace*{1em}{\it Case 2.2}\,: 
\,$s\ _{\neq}\fleche{u',v'}_G\ s'\ \fleche{u'',v''}_G$ \,for some \,$s'$ \,with 
\,$u = u'u''$ \,and $v = v'v''$.\\
\hspace*{2em}By hypothesis \,$t\ \fleche{u',v'}_G\ t'$ \,for some vertex 
\,$t'$.\\
\hspace*{2em}By induction hypothesis \,$t'\ \fleche{u'',v''}_G$ \,hence 
\,$t\ \fleche{u,v}_G$\,.\\[0.25em]
{\bf ii)} Let us check (b).\\
$\Longleftarrow$\,: \,by (a) for \,$G$ \,source-complete and 
elementary-propagating.\\
$\Longrightarrow$\,: \,Let \,$G$ \,be a deterministic and propagating graph.\\
Let us check that \,$G$ \,is elementary-propagating.\\
Let \,$s\ _{\neq}\fleche{u,v}_G$ \,with \,$u,v \in A_G^*$ \,and let 
\,$t \in V_G$\,. 
So \,$t\ \fleche{u,v}_G$ \,and \,$uv \neq \varepsilon$.\\
As \,$G$ \,is deterministic and for \,$u,v \neq \varepsilon$, the words \,$u$ 
\,and \,$v$ \,have not the same first letter.\\
As yet done for (a), there exists \,$u'u'' = u$ \,and \,$v'v'' = v$ \,such that 
\,$t\ _{\neq}\fleche{u',v'}_G\,\fleche{u'',v''}_G$\,.\\
So \,$s\ \fleche{u',v'}_G$\,. 
For \,$G$ \,deterministic, $u'' = v'' = \varepsilon$ \,hence 
\,$t\ _{\neq}\fleche{u,v}_G$\,.\\[0.25em]
{\bf iii)} (c) \,follows from Lemma~\ref{ForwardTransitive}.\\[0.25em]
{\bf iv)} (d) \,follows from (c) and Lemmas~\ref{AccessTransitive} and 
\ref{ForwardTransitive}.\\[0.25em]
{\bf v)} Let us check (e). Let \,$r$ \,be a propagating root of a 
deterministic and target-complete graph \,$G$. 
We have to show that \,$G$ \,is propagating.\\
Let \,$s\ \fleche{u,v}_G$ \,with \,$u,v \in A_G^*$. 
We have to check that \,$r\ \fleche{u,v}_G$\,.\\
As \,$r$ \,is a root, $r\ \fleche{w}_G\ s$ \,for some \,$w \in A_G^*$.\\
As \,$G$ \,is target complete, there exists a vertex \,$r_0$ \,such 
that \,$r_0\ \fleche{w}_G\ r$.\\
As \,$r\ \fleche{wu,wv}_{\!\!G}$ \,and \,$r$ \,is propagating, 
$r_0\  \fleche{wu,wv}_{\!\!G}$\,. 
As \,$G$ \,is deterministic, $r\ \fleche{u,v}_G$\,.
\qed\\[1em]
Note that the determinism condition in 
Proposition~\ref{ForwardVertexTransitive}~(b) and (c) is necessary. 
For instance the following propagating and non deterministic graph:
\begin{center}
\includegraphics{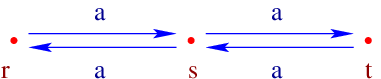}\\
\end{center}
is not elementary-propagating (and not forward vertex-transitive) since 
\,$s\ _{\neq}\!\fleche{aa,aa}$ \,but \,$\neg(r\ _{\neq}\!\fleche{aa,aa})$.\\
Note also that the rooted condition in 
Proposition~\ref{ForwardVertexTransitive}~(e) is necessary. 
For instance, the inverse semiline 
\,$\{\ n+1\ \fleche{a}\ n\ |\ n \in \entier\ \}$ \,is deterministic, 
target-complete, and is not a propagating graph while $0$ is a propagating 
co-root.\\[0.5em]
Let us generalize to the chains the path commutation in a graph.\\
A vertex \,$r$ \,is a {\it chain-commutative vertex} \,for a graph \,$G$ \,if 
\,$r$ \,is commutative for \,$\overline{G}$ \,{\it i.e.}\\
\hspace*{0.5em}$r\ \fleche{u}_G\ s \ \Longrightarrow \ r\ \fleche{v}_G\ s$ 
\,for any \,$s \in V_G$ \,and any \,$u,v \in (A_G \cup \overline{A_G})^*$ 
\,such that \,$u \,\cong_A \,v$\,.\\
We say that \,$G$ \,is a {\it chain-commutative graph} \,if \,$\overline{G}$ 
\,is a commutative graph \,{\it i.e.} \,all the vertices are 
chain-commutative for \,$G$.\\
As in Lemma~\ref{Commutative}, we can express locally the chain-commutation. 
A vertex \,$r$ \,is a {\it locally chain-commutative vertex} \,for a graph 
\,$G$ \,if \,$r$ \,is locally commutative for \,$\overline{G}$ \,{\it i.e.} 
\,$r\ \fleche{ab}_G\ s \ \Longrightarrow \ r\ \fleche{ba}_G\ s$ 
\,for any \,$s \in V_G$ \,and \,$a,b \in A_G \cup \overline{A_G}$\,. 
Let us apply Lemma~\ref{Commutative}.
\begin{corollary}\label{ChainCommutative}
A graph is chain-commutative \,{\rm iff} \,all its vertices are locally 
chain-commutative.
\end{corollary}
Any chain-commutative graph is commutative and the converse is true when the 
graph is deterministic, co-deterministic, source and target-complete.
\begin{lemma}\label{Chain-Commutative}
Let \,$G$ \,be a deterministic, co-deterministic, source and target-complete 
graph.\\
\hspace*{6.5em}If \,$G$ \,is commutative then \,$G$ \,is chain-commutative.
\end{lemma}
\proof\mbox{}\\
Let \,$s\ \fleche{ab}_G\ t$ \,where \,$a,b \in A_G \cup \overline{A_G}$\,. 
By Corollary~\ref{ChainCommutative}, it remains to check that 
\,$s\ \fleche{ba}_G\ t$.\\
We distinguish the four complementary cases below.\\
{\it Case 1}\,: $a,b \in A_G$\,. As \,$s$ \,is locally commutative, 
$s\ \fleche{ba}_G\ t$.\\
{\it Case 2}\,: $\overline{a},\overline{b} \in A_G$\,. 
So \,$t\ \fleche{\overline{b}\,\overline{a}}_G\ s$. 
As \,$t$ \,is locally commutative, 
$t\ \fleche{\overline{a}\,\overline{b}}_G\ s$ \,{\it i.e.} 
\,$s\ \fleche{ba}_G\ t$.\\
{\it Case 3}\,: $\overline{a},b \in A_G$\,. 
So \,$s\ \inverse{\overline{a}}_G\ r\ \fleche{b}_G\ t$ \,for some 
vertex \,$r$.\\
\hspace*{1em}As \,$G$ \,is source-complete, $s\ \fleche{b}_G\ r'$ \,for some 
vertex \,$r'$. So \,$r\ \fleche{\overline{a}b}_G\ r'$.\\
\hspace*{1em}As \,$r$ \,is locally commutative, 
$r\ \fleche{b\overline{a}}_G\ r'$.\\
\hspace*{1em}As \,$G$ \,is deterministic, \,$t\ \fleche{\overline{a}}_G\ r'$ 
\,hence \,$s\ \fleche{ba}_G\ t$.\\
{\it Case 4}\,: $a,\overline{b} \in A_G$\,. 
So \,$s\ \fleche{a}_G\ r\ \inverse{\overline{b}}_G\ t$ \,for some vertex \,$r$.
\\
\hspace*{1em}As \,$G$ \,is target-complete, $r'\ \fleche{a}_G\ t$ \,for some 
vertex \,$r'$. So \,$r'\ \fleche{a\overline{b}}_G\ r$.\\
\hspace*{1em}As \,$r'$ \,is locally commutative, 
$r'\ \fleche{\overline{b}a}_G\ r$.\\
\hspace*{1em}As \,$G$ \,is co-deterministic, 
\,$r'\ \fleche{\overline{b}}_G\ s$ \,hence \,$s\ \fleche{ba}_G\ t$.
\qed\mbox{}\\[1em]
Let us apply Lemmas~\ref{CommutativeForward} and \ref{Chain-Commutative}.
\begin{lemma}\label{PropagatingChainCommutative}
Let \,$G$ \,be a deterministic, co-deterministic, source and target-complete 
graph.\\
\hspace*{5.2em}If \,$G$ \,has a propagating locally commutative vertex then 
\,$G$ \,is chain-commutative.
\end{lemma}
Let us generalize to the chains the path propagation in a graph. Note that\\
\hspace*{6em}$s\ \fleche{u,v}_G \ \ \Longleftrightarrow \ \ 
s\ \fleche{u\,\widetilde{\overline{v}}}_G\ s$ \ \ for any \,$s \in V_G$ \,and 
\,$u,v \in (A_G \cup \overline{A_G})^*$.\\
A vertex \,$r$ \,is a {\it chain-propagating vertex} \,for a graph \,$G$ \,if 
\,$r$ \,is propagating for \,$\overline{G}$ \,{\it i.e.}\\
\hspace*{3em}for any \,$u \in (A_G \cup \overline{A_G})^*$, \ \ 
$r\ \fleche{\varepsilon,u}_G \ \ \ \Longrightarrow \ \ \ \forall\ s \in V_G\ 
(s\ \fleche{\varepsilon,u}_G\,)$.\\
Let us apply Lemma~\ref{ForwardTransitive}.
\begin{corollary}\label{Transitive}
For any deterministic and co-deterministic graph \,$G$ \,and any vertices 
\,$s,t$,\\
\hspace*{9em}$s \,\simeq_G t \ \ \Longleftrightarrow \ \ 
\forall\ u \in (A_G \cup \overline{A_G})^*,\,\ (s\ \fleche{\varepsilon,u}_G \ 
\Longleftrightarrow \ t\ \fleche{\varepsilon,u}_G\,)$.
\end{corollary}
A {\it chain-propagating graph} \,is a graph whose all its vertices are 
chain-propagating. 
The line \,$\{\ n\ \fleche{a}\ n+1\ |\ n \in \relatif\ \}$ \,is 
chain-propagating but the semiline 
\,$\{\ n\ \fleche{a}\ n+1\ |\ n \in \entier\ \}$ \,is not chain-propagating 
since \,$\neg(0\ \fleche{\overline{a}})$. 
We denote by \,$C_G(s) \,= \,\{\ u \in (A_G \cup \overline{A_G})^*\ |\ 
s\ _{\neq}\fleche{\varepsilon,u}_{\overline{G}}\ \}$ \,the language of non empty 
words labeling the elementary cycles from \,$s$\,.\\
Let us apply Proposition~\ref{ForwardVertexTransitive}.
\begin{corollary}\label{VertexTransitive}
{\bf a)} A source and target-complete graph \,$G$ \,is chain-propagating \ 
{\rm if and only~if}\\
\hspace*{6em}$s\ _{\neq}\fleche{\varepsilon,u}_G \ \ \Longrightarrow \ \ 
t\ \fleche{\varepsilon,u}_G\,$ \ for any \,$s,t \in V_G$ \,and 
\,$u \in (A_G \cup \overline{A_G})^*$\,.\\
\hspace*{0em}{\bf b)} Let \,$G$ \,be a deterministic and co-deterministic, 
source and target-complete graph,\\
\hspace*{6em}$G$ \,is chain-propagating \ \ $\Longleftrightarrow \ \ 
C_G(s) = C_G(t)$ \,for any \,$s,t \in V_G$\,.\\
\hspace*{0em}{\bf c)} A deterministic and co-deterministic graph is 
chain-propagating \,{\rm iff} \,it is vertex transitive.\\
\hspace*{0em}{\bf d)} Let \,$G$ \,be a connected deterministic and 
co-deterministic graph and \,$r$ \,be a vertex,\\
\hspace*{0.5em}$G$ \,is chain-propagating \ {\rm iff} \ 
$r\ \fleche{\varepsilon,u}_G \ \Longleftrightarrow \ s\ 
\fleche{\varepsilon,u}_G\,$ \,for any \,$r\ \deriv{}_G\ s$ \,and 
\,$u \in (A_G \cup \overline{A_G})^*$.\\
\hspace*{0em}{\bf e)} For any connected, deterministic and co-deterministic 
graph \,$G$,\\
\hspace*{0.5em}$G$ \,is vertex-transitive \ {\rm iff} \ $G$ \,is source and 
target-complete with a chain-propagating vertex.
\end{corollary}
Note that the connected condition in Corollary~\ref{VertexTransitive}~(e) is 
necessary. For instance, the non-connected graph 
\,$\{\ n\ \fleche{a}\ n+1\ |\ n \in \relatif\ \} \,\cup 
\,\{\ \omega\ \fleche{a}\ \omega\ \}$ \,is not vertex-transitive but it is 
deterministic, co-deterministic, source and target-complete, and any 
\,$n \in \relatif$ \,is a chain-propagating vertex.

\section{Generalized Cayley graphs of magmas}\label{CayMag}

{\indent}We present graph-theoretic characterizations for the generalized 
Cayley graphs of magmas with a left identity, and with an identity 
(see Theorem~\ref{MagmaLeftIdentity}). 
These characterizations are then refined to the commutative magmas with an 
identity (see Theorem~\ref{MagmaCommutativity}).\\[-0.5em]

A {\it magma} \,is a set \,$M$ \,equipped with a binary operation 
\,$\cdot\,: M{\croix}M\ \fleche{}\ M$ \,that sends any two elements 
\,$p,q \in M$ \,to the element \,$p \cdot q$.\\
Given a subset \,$Q \subseteq M$ \,and an injective mapping 
\,$\inter{\ }\,: Q\ \fleche{}\ A$, we define the graph\\[0.25em]
\hspace*{10em}
${\cal C}\inter{M,Q}\ =\ \{\ p\ \fleche{\interFootnote{q}}\ p \cdot q\ |\ 
p \in M \,\wedge \,q \in Q\ \}$\\[0.25em]
which is called a {\it generalized Cayley graph} \,of \,$M$. 
It is of  vertex set \,$M$ \,and of label set 
\,$\inter{Q} \,= \,\{\ \inter{q}\ |\ q \in Q\ \}$. 
We denote \,${\cal C}\inter{M,Q}$ \,by \,${\cal C}(M,Q)$ \,when 
\,$\inter{\ }$ \,is the identity. 
We also write \,${\cal C}\inter{M}$ \,instead of \,${\cal C}\inter{M,M}$ \,and 
\,${\cal C}(M) \,= \,{\cal C}(M,M) \,= \,\{\ p\ \fleche{q}\ p \cdot q\ |\ 
p,q \in M\ \}$.\\
Let us give basic properties of these generalized Cayley graphs.
\begin{fact}\label{Magma}
Any generalized Cayley graph of a magma is deterministic and source-complete.
\end{fact}
A magma \,$(M,\cdot)$ \,is {\it left-cancellative} \,if 
\,$r \cdot p = r \cdot q \ \Longrightarrow \ p = q$ \,for any \,$p,q,r \in M$.\\
Similarly \,$(M,\cdot)$ \,is {\it right-cancellative} \,if 
\,$p \cdot r = q \cdot r \ \Longrightarrow \ p = q$ \,for any \,$p,q,r \in M$.\\
A magma is {\it cancellative} \,if it is both left-cancellative and 
right-cancellative.
\begin{fact}\label{CancelMagma}
Any generalized Cayley graph of a left-cancellative magma is simple.\\
\hspace*{4.8em}Any generalized Cayley graph of a right-cancellative magma is 
co-deterministic.
\end{fact}
Recall that an element \,$e$ \,of \,$M$ \,is a {\it left identity} (resp. 
{\it right identity}) \,if \,$e{\cdot}p = p$ (resp. \,$p{\cdot}e = p$) for any 
\,$p \in M$. 
\begin{fact}\label{1Root}
Any left identity of a magma \,$M$ \,is an out-simple 1-root of \,${\cal C}(M)$.
\end{fact}
If \,$M$ \,has a left identity \,$e$ \,and a right identity \,$e'$ \,then 
\,$e = e \cdot e' = e'$ \,is the {\it identity} \,of \,$M$ \,which is called a 
{\it unital magma}.
\begin{fact}\label{LoopPropagating}
The identity of any unital magma \,$M$ \,is loop-propagating for 
\,${\cal C}(M)$.
\end{fact}
\proof\mbox{}\\
Assume that \,$M$ \,has an identity \,$1$ \,with an \,$a$-loop on 
\,${\cal C}(M)$\,: \,$1\ \fleche{a}_{{\cal C}(M)}\ 1$ \,with \,$a \in A$.\\
By definition of \,${\cal C}(M)$, we have \,$1{\cdot}a = 1$ \,{\it i.e.} 
\,$a = 1$.\\
For any vertex \,$s$, we have \,$s\ \fleche{a}_{{\cal C}(M)}\ s{\cdot}a \,= 
\,s{\cdot}1 \,= \,s$.
\qed\\[1em]
Recall that a magma \,$M$ \,is {\it commutative} \,if 
\,$p \cdot q = q \cdot p$ \,for any \,$p,q \in M$.
\begin{fact}\label{CommutativeMagma}
Any left identity of a commutative magma is locally commutative for 
\,${\cal C}(M)$.
\end{fact}
We say that (the operation \,$\cdot$ \,of) a magma \,$M$ \,is 
{\it left-invertible} (resp. {\it right-invertible}) with respect to 
\,$e \in M$ \,if for any \,$p \in M$ \,there exists 
\,$\overline{p} \in M$ \,such that \,$\overline{p} \cdot p = e$ (resp. 
\,$p \cdot \overline{p} = e$).\\
A unital magma is left (resp. right)-invertible if it is w.r.t. its unit 
element.
\begin{fact}\label{InvertibleMagma}
For any left-invertible $($resp. right-invertible$)$ magma \,$M$ w.r.t. \,$e$,\\
\hspace*{5.5em}$e$ \,is a target-complete vertex $($resp. is an $1$-coroot$)$ 
for \,${\cal C}(M)$.
\end{fact}
\proof\mbox{}\\
For any \,$p \in M$, \,$\overline{p}\ \fleche{p}_{{\cal C}(M)}\ e$ \,for 
\,$M$ \,left-invertible, and \,$p\ \fleche{\overline{p}}_{{\cal C}(M)}\ e$ \,for 
\,$M$ \,right-invertible.
\qed\\[1em]
The previous facts are basic properties of the generalized Cayley graphs of 
magmas with a left-identity that characterize them.
\begin{lemma}\label{EdgeProduct}
Let \,$r$ \,be an $1$-propagating $1$-root of a deterministic graph \,$G$.\\
\hspace*{1em}We can define the {\rm edge-operation} \,$\croix_r$ \,for any 
\,$s,t \in V_G$ \,by\\
\hspace*{10em}$s\ \fleche{a}_G\ s \,\croix_r \,t$ \ if \ $r\ \fleche{a}_G\ t$ 
\,for some \,$a \in A$.\\
\hspace*{1em}Then \,$(V_G,\croix_r)$ \,is a magma of left identity \,$r$.\\
\hspace*{1em}If \,$r$ \,is loop-propagating then \,$r$ \,is an identity.\\
\hspace*{1em}If \,$r$ \,is locally commutative then \,$\croix_r$ \,is 
commutative.\\
\hspace*{1em}If \,$G$ \,is $1$-propagating \,then \,$\croix_r$ \,is 
left-cancellative.\\
\hspace*{1em}If \,$G$ \,is co-deterministic \,then \,$\croix_r$ \,is 
right-cancellative.\\
\hspace*{1em}If \,$r$ \,is target-complete \,then \,$\croix_r$ \,is 
left-invertible w.r.t. \,$r$.\\
\hspace*{1em}If \,$r$ \,is a source-complete \,$1$-coroot \,then 
\,$\croix_r$ \,is right-invertible w.r.t. \,$r$.\\
\hspace*{1em}For any \,$B \subseteq A_G$ \,such that \,$r$ \,is 
source-complete and out-simple for $G^{|B}$,\\
\hspace*{1em}$G^{|B} \,= \,{\cal C}\inter{V_G,Q}$ \,for 
\,$Q = \{\,q\,|\,\exists\ b \in B\ (r\ \fleche{b}_G\ q)\,\}$ \,with 
\,$\inter{q} \,= \,b$ \,for \,$b \in B$ \,and~\,$r\ \fleche{b}_G\ q$.
\end{lemma}
\proof\mbox{}\\
{\bf i)} Let \,$s,t \in V_G$\,. 
\,Let us check that \,$s \,\croix_r \,t$ \,is well-defined.\\
As \,$r$ \,is an \,$1$-root, there exists \,$a \in A$ \,such that 
\,$r\ \fleche{a}_G\ t$.\\
As \,$r$ \,is \,$1$-propagating, there exists \,$x$ \,such that 
\,$s\ \fleche{a}_G\ x$.\\
Let \,$r\ \fleche{b}_G\ t$. So \,$r\ \fleche{a,b}_G$\,. 
As \,$r$ \,is $1$-propagating, we have \,$s\ \fleche{a,b}_G$\,.\\
As \,$G$ \,is deterministic, we get \,$s\ \fleche{b}_G\ x$. 
Thus \,$s \,\croix_r \,t\ =\ x$.\\
The operation \,$\croix_r$ \,is illustrated as follows:
\begin{center}
\includegraphics{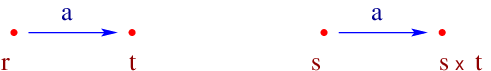}\\
\end{center}
We denote \,$\croix_r$ \,by \,$\cdot$ \,in the rest of this 
proof.\\[0.25em]
{\bf ii)} Let us check that \,$r$ \,is a left identity of \,$(V_G,\cdot)$.\\
Let \,$s \in V_G$\,. As \,$r$ \,is an \,$1$-root, there exists \,$a \in A$ 
\,such that \,$r\ \fleche{a}_G\ s$.\\
By definition of \,$\cdot$ \,we have \,$r\ \fleche{a}_G\ r{\cdot}s$. 
As \,$G$ \,is deterministic, we get \,$r{\cdot}s = s$.\\[0.25em]
{\bf iii)} Assume that \,$r$ \,is loop-propagating. 
Let us check that \,$r$ \,is also a right identity.\\
As \,$r$ \,is an \,$1$-root, there is \,$a \in A_G$ \,such that 
\,$r\ \fleche{a}_G\ r$.\\
Let \,$s \in V_G$\,. \,As \,$r$ \,is loop-propagating, we get 
\,$s\ \fleche{a}_G\ s$.\\
By definition of \,$\cdot$ \,we have \,$s\ \fleche{a}\ s{\cdot}r$.\\
As \,$G$ \,is deterministic, \,$s = s{\cdot}r$. 
Thus \,$r$ \,is a right identity.\\[0.25em]
{\bf iv)} Assume that \,$r$ \,is locally commutative. 
Let us check that \,$\cdot$ \,is commutative.\\
Let \,$s,t \in V_G$\,. 
There exists \,$a,b \in A_G$ \,such that \,$r\ \fleche{a}\ s$ \,and 
$r\ \fleche{b}\ t$.\\
So \ $s\ \fleche{b}\ s{\cdot}t$ \ and \ $t\ \fleche{a}\ t{\cdot}s$. 
Thus \,$r\ \fleche{ab}\ s{\cdot}t$ \ and \ $r\ \fleche{ba}\ t{\cdot}s$.\\
As \,$r$ \,is locally commutative, $r\ \fleche{ba}\ s{\cdot}t$. 
As \,$G$ \,is deterministic, we get \,$s{\cdot}t = t{\cdot}s$.\\[0.25em]
{\bf v)} Assume that \,$G$ \,is \,$1$-propagating. 
Let us check that \,$\cdot$ \,is left-cancellative.\\
Assume that \,$s{\cdot}t \,= \,s{\cdot}t'$.\\
As \,$r$ \,is an \,$1$-root, there exists \,$a,a' \in A_G$ \,such that 
\,$r\ \fleche{a}_G\ t$ \,and \,$r\ \fleche{a'}_G\ t'$.\\
By definition of \,$\cdot$ \,we get \,$s\ \fleche{a}_G\ s \cdot t$ \,and 
\,$s\ \fleche{a'}_G\ s{\cdot}t' \,= \,s{\cdot}t$. 
Thus \,$s\ \fleche{a,a'}_G$\,.\\
As \,$G$ \,is \,$1$-propagating, $r\ \fleche{a,a'}_G$\,. 
As \,$G$ \,is deterministic, it follows that \,$t = t'$.\\[0.25em]
{\bf vi)} Assume that \,$G$ \,is co-deterministic.
Let us check that \,$\cdot$ \,is right-cancellative.\\
Let \,$s,s',t \in V_G$ \,such that \,$s{\cdot}t \,= \,s'{\cdot}t$.\\
There exists \,$a \in A_G$ \,such that \,$r\ \fleche{a}\ t$. 
So \ $s\ \fleche{a}\ s{\cdot}t$ \ and \ 
$s'\ \fleche{a}\ s'{\cdot}t \,= \,s{\cdot}t$.\\
As \,$G$ \,is co-deterministic, we get \,$s = s'$.\\[0.25em]
{\bf vii)} Assume that \,$r$ \,is target-complete. 
Let us check that \,$\cdot$ \,is left-invertible w.r.t. \,$r$.\\
Let \,$s \in V_G$\,. 
As \,$r$ \,is an \,$1$-root, there exists \,$a \in A_G$ \,such that 
\,$r\ \fleche{a}_G\ s$.\\
As \,$r$ \,is target-complete, there exists \,$\overline{s} \in V_G$ \,such 
that \,$\overline{s}\ \fleche{a}_G\ r$.\\
So \,$\overline{s}\ \fleche{a}_G\ \overline{s} \cdot s$. 
As \,$G$ \,is deterministic, \,$\overline{s} \cdot s = r$.\\[0.25em]
{\bf viii)} Assume that \,$r$ \,is source-complete and is an $1$-coroot.\\
Let us check that \,$\cdot$ \,is right-invertible w.r.t. \,$r$.\\
Let \,$s \in V_G$\,. 
As \,$r$ \,is an $1$-coroot, there exists \,$a \in A_G$ \,such that 
\,$s\ \fleche{a}_G\ r$.\\
As \,$r$ \,is source-complete, there exists \,$\overline{s} \in V_G$ \,such 
that \,$r\ \fleche{a}_G\ \overline{s}$.\\
So \,$s\ \fleche{a}_G\ s \cdot \overline{s}$. 
As \,$G$ \,is deterministic, \,$s \cdot \overline{s} = r$.\\[0.25em]
{\bf ix)} Let \,$B \subseteq A_G$ \,such that \,$r$ \,is out-simple and 
source-complete for \,$H = G^{|B}$.\\
Let \,\,$Q \,= \,\{\ q\ |\ \exists\ b \in B\ (r\ \fleche{b}_G\ q)\ \} \,= 
\,\{\ q\ |\ \exists\ b \in A\ (r\ \fleche{b}_H\ q)\ \}$.\\
As \,$r$ \,is out-simple for \,$H$, we can define the mapping 
\,$\inter{\ } : Q \,\fleche{} \,B$ \,by \,$\inter{q} \,= \,b$ \,for 
\,$r\ \fleche{b}_H\ q$.\\
As \,$H$ \,is deterministic, \,$\inter{\ }$ \,is an injection. 
As \,$r$ \,is source-complete for \,$H$, \,$\inter{\ }$ \,is a bijection.\\
Let us show that \,$H \,= \,{\cal C}\inter{V_G,Q}$.\\
$\subseteq$\,: \,Let \,$s\ \fleche{b}_H\ t$.
As \,$r$ \,is source-complete for \,$H$, there exists \,$q$ \,such that 
\,$r\ \fleche{b}_H\ q$.\\
So \,$s\ \fleche{b}_G\ s{\cdot}q$. 
As \,$G$ \,is deterministic, $s{\cdot}q \,= \,t$. 
As \,$\inter{q} \,= \,b$, we get 
\,$s\ \fleche{b}_{{\cal C}\interInd{V_G,Q}}\ s{\cdot}q \,= \,t$.\\[0.25em]
$\supseteq$\,: \,Let \,$s\ \fleche{b}_{{\cal C}\interInd{V_G,Q}}\ t$.
There exists \,$q \in Q$ \,such that \,$\inter{q} = b$ \,and 
\,$s{\cdot}q \,= \,t$.\\
Thus \,$b \in B$ \,and \,$r\ \fleche{b}_G\ q$. 
So \,$s\ \fleche{b}_H\ s{\cdot}q \,= \,t$.
\qed\\[1em]
We get a fully graph-theoretic characterization for the Cayley graphs 
\,${\cal C}\inter{M}$ \,of any magma~\,$M$ \,with a left identity.
\begin{proposition}\label{CayleyMagma}
A graph \,$G$ \,is equal to \,${\cal C}\inter{M}$ \,for some magma \,$M$ 
\,with a left identity~\,$r$ \,$($resp. commutative magma, with an identity, 
left-cancellative, right-cancellative, left-invertible w.r.t. $r$, 
right-invertible w.r.t. $r)$ \ {\rm if and only if}\\
$G$ \,is a deterministic and source-complete graph and \,$r$ \,is an out-simple 
\,$1$-root $($and resp. $r$ \,is locally commutative, 
$r$ \,is loop-propagating, $G$ \,is simple, $G$ \,is co-deterministic, 
$r$ \,is target-complete, $r$ is an \,$1$-coroot\,$)$.
\end{proposition}
\proof\mbox{}\\
$\Longrightarrow$\,: \,Let \,$G \,= \,{\cal C}\inter{M}$ \,for some 
magma \,$(M,\cdot)$ \,with a left identity \,$r$, and some 
injective mapping \,$\inter{\ }$. 
By Fact~\ref{Magma}, \,$G$ \,is deterministic and source-complete.\\ 
By Fact~\ref{1Root}, $r$ \,is out-simple and is an \,$1$-root of \,$G$.\\
If \,$M$ \,is commutative then by Fact~\ref{CommutativeMagma}, $r$ \,is 
locally commutative.\\
If \,$r$ \,is an identity then by Fact~\ref{LoopPropagating}, $r$ \,is 
loop-propagating.\\
If  \,$\cdot$ \,is left-cancellative (resp. right-cancellative) then by 
Fact~\ref{CancelMagma}, \,$G$ \,is simple (resp. co-deterministic).\\
If \,$M$ \,is left-invertible (resp. right-invertible) w.r.t. \,$r$ \,then by 
Fact~\ref{InvertibleMagma}, $r$ \,is target-complete (resp. is an $1$-coroot).
\\[0.25em]
$\Longleftarrow$\,: \,By Fact~\ref{SourcePropagating} and by 
Lemma~\ref{EdgeProduct} (resp. by Fact~\ref{SourceCompletePropagating}).
\qed\\[1em]
We can now present a graph-theoretic characterization for the generalized 
Cayley graphs of magmas with a left identity, or with an identity. 
\begin{theorem}\label{MagmaLeftIdentity}
A graph is a generalized Cayley graph of a magma with a 
left identity
$($resp. an identity$)$
 \ {\rm if and only if} \ it is 
deterministic, source-complete with an out-simple $($resp. and 
loop-propagating$)$
vertex.
\end{theorem}
\proof\mbox{}\\
$\Longrightarrow$\,: \,By Facts~\ref{Magma}, \ref{1Root}, 
(resp. \ref{LoopPropagating}), 
or by Proposition~\ref{CayleyMagma} having 
\,${\cal C}\inter{M,Q} \,= \,{\cal C}\inter{M}^{|\interInd{Q}}$.\\
$\Longleftarrow$\,: \,Let \,$G$ \,be a deterministic and source-complete graph 
with an out-simple vertex \,$r$.\\
Let \,$Q \,= \,\{\ q\ |\ r\ \fleche{}_G\ q\ \}$ \,be the set of successors of 
\,$r$. We define the graph\\[0.25em]
\hspace*{3em}\begin{tabular}{rcl}
$G_1$ & $=$ & $\{\ s\ \fleche{q}\ t\ |\ \exists\ a\ (s\ \fleche{a}_G\ t 
\,\wedge\ r\ \fleche{a}_G\ q)\ \} \,\cup 
\,\{\ r\ \fleche{s}\ s\ |\ s \in V_G-Q\,\}$\\[0.25em]
 & $\cup$ & $\{\ s\ \fleche{t}\ r\ |\ s \in V_G-\{r\} \,\wedge 
\,t \in V_G - Q\ \}$.
\end{tabular}\\[0.25em]
Here is an example with \,$r \not\in Q$ \,loop-propagating:
\begin{center}
\includegraphics{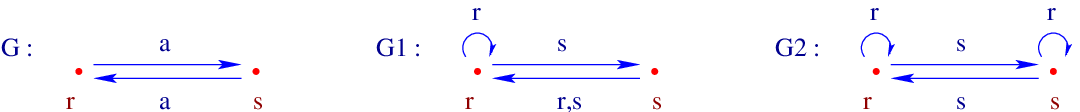}\\
\end{center}
and another example with \,$r \in Q$ \,not loop-propagating:
\begin{center}
\includegraphics{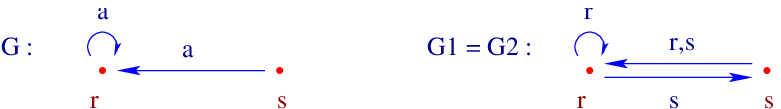}\\
\end{center}
Thus \,$G_1$ \,remains deterministic and source-complete with 
\,$V_{G_1} \,= \,V_G \,= \,A_{G_1}$\,.\\
As \,$r\ \fleche{s}_{G_1}\ s$ \,for any \,$s \in V_G$\,, \,$r$ \,is an 
out-simple $1$-root of \,$G_1$.\\
By Fact~\ref{SourcePropagating}, $r$ \,is $1$-propagating.\\
By Lemma~\ref{EdgeProduct} applied to $G_1$ and $r$, 
$(V_G,\croix_r)$ is a magma of left identity \,$r$ \,and 
$G_1={\cal C}(V_G)$~{\it i.e.}\\
\hspace*{10em}$s\ \fleche{t}_{G_1}\ s \,\croix_r \,t$ \ for any 
\,$s,t \in V_G$\,.\\[0.25em]
Thus \,$G \,= \,{\cal C}\inter{V_G}^{|\interInd{Q}} \,= 
\,{\cal C}\inter{V_G\,,\,Q}$ \,with \,$\inter{q} = a$ \,for any 
\,$r\ \fleche{a}_G\ q$.\\
Note that for \,$r \not\in Q$ \,and \,$s \neq r$ (as for the first example), 
$r$ \,is not loop-propagating for \,$G_1$ \,since \,$r\ \fleche{r}_{G_1}\ r$ 
\,and \,$s\ \fleche{r}_{G_1}\ r$. 
Thus we refine \,$G_1$ \,in the following graph:\\[0.25em]
\hspace*{0em}\begin{tabular}{rcl}
$G_2$ & $=$ & $\{\ s\ \fleche{q}\ t\ |\ \exists\ a\ (s\ \fleche{a}_G\ t 
\,\wedge\ r\ \fleche{a}_G\ q)\ \} \,\cup 
\,\{\ r\ \fleche{s}\ s\ |\ s \in V_G-Q\,\}$\\[0.25em]
& $\cup$ & $\{\ s\ \fleche{r}\ s\ |\ s \in V_G-\{r\} \,\wedge \,r \not\in Q\ \}
\ \cup\ \{\ s\ \fleche{t}\ r\ |\ s \in V_G-\{r\} \,\wedge \,t \in V_G - (Q \cup \{r\})
\ \}$.
\end{tabular}\\[0.25em]
Note that \,$G_2$ \,remains deterministic and source-complete with 
\,$V_{G_2} \,= \,V_G \,= \,A_{G_2}$\,.\\
As \,$r\ \fleche{s}_{G_2}\ s$ \,for any \,$s \in V_G$\,, \,$r$ \,is an 
out-simple $1$-root of \,$G_2$.\\
Assume that \,$r$ \,is loop-propagating for \,$G$. 
In particular \,$G_1 = G_2$ \,for \,$r \in Q$.\\
Let us check that \,$r$ \,remains loop-propagating for \,$G_2$. 
Let \,$r\ \fleche{s}_{G_2}\ r$ \,and \,$t \in V_G$\,.\\
\hspace*{1em}Either \,$s \not\in Q$. Thus \,$s = r$. So \,$r \not\in Q$ \,hence 
\,$t\ \fleche{r}_{G_2}\ t$.\\
\hspace*{1em}Or \,$s \in Q$ \,{\it i.e.} \,$r\ \fleche{a}_G\ s$ \,for some 
\,$a \in A$. 
So  \,$r\ \fleche{s}_{G_2}\ s$.\\
\hspace*{1em}As \,$G_2$ \,is deterministic, $s = r$. 
Thus \,$r\ \fleche{a}_G\ r$.\\
\hspace*{1em}As \,$r$ \,is loop-propagating for \,$G$, we get 
\,$t\ \fleche{a}_G\ t$ \,hence \,$t\ \fleche{s}_{G_2}\ t$.\\
By Lemma~\ref{EdgeProduct}, $(V_G,\croix_r)$ \,is a magma of identity 
\,$r$ \,with \,$G_2 \,= \,{\cal C}(V_G)$.\\
Thus \,$G \,= \,{\cal C}\inter{V_G}^{|\interInd{Q}} \,= 
\,{\cal C}\inter{V_G\,,\,Q}$ \,with \,$\inter{q} = a$ \,for any 
\,$r\ \fleche{a}_G\ q$.
\qed\\[1em]
For instance consider the magma \,$(\relatif,\cdot)$ \,where 
\,$m \cdot n = -m + n$ \,for any \,$m,n \in \relatif$. 
There is no right identity and \,$0$ \,is the unique left identity. 
The generalized Cayley graph \,$G \,= \,{\cal C}\inter{\relatif,\{1,2\}}$ 
\,with \,$\inter{1} = a$ \,and \,$\inter{2} = b$ \,is the following simple 
graph:
\begin{center}
\includegraphics{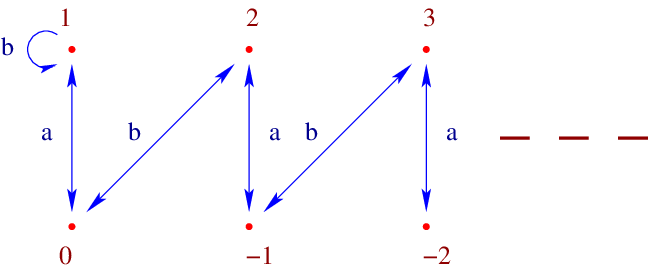}\\
\end{center}
Any vertex other than $1$ is loop-propagating. 
By Theorem~\ref{MagmaLeftIdentity}, $G$ \,is a generalized Cayley graph of a 
unital magma. Precisely by Lemma~\ref{EdgeProduct}, 
$G \,= \,{\cal C}\inter{\relatif,\{1,2\}}$ \,with \,$\inter{1} = a$ \,and 
\,$\inter{2} = b$ \,for the unital magma \,$(\relatif,\croix_0)$ \,defined for 
any \,$n$ \,by \,$0 \,\croix_0 \,n = n$ \,and~for~any~\,$m \neq 0$~\,by
\\[0.25em]
\hspace*{10em}{$m \,\croix_0 \,n \ = \ \left\{\begin{tabular}{ll}
$m$ & if \ $n = 0$\\[0.25em]
$-m + n$ & if \ $n = 1,2$\\[0.25em]
$0$ & otherwise.
\end{tabular}\right.$}\\[0.25em]
We strengthen Theorem~\ref{MagmaLeftIdentity} for any commutative magma with a 
unit element.
\begin{theorem}\label{MagmaCommutativity}
A graph is a generalized Cayley graph of a commutative unital magma \ 
{\rm iff}\\
\hspace*{7.3em}it is deterministic, source-complete with an out-simple, 
loop-propagating\\
\hspace*{7.3em}and locally commutative
vertex.
\end{theorem}
\proof\mbox{}\\
$\Longrightarrow$\,: \,By Facts~\ref{Magma}, \ref{1Root}, 
\ref{LoopPropagating}, \ref{CommutativeMagma}.\\
$\Longleftarrow$\,: \,Let \,$G$ \,be a deterministic and source-complete 
graph.\\ 
Let \,$r$ \,be an out-simple, loop-propagating and locally commutative vertex.\\
Let \,$Q \,= \,\{\ q\ |\ r\ \fleche{}_G\ q\ \}$ \,and 
\,$Q_r = Q \cup \{r\}$.\\
The graph \,$G_2$ \,defined in the proof of Theorem~\ref{MagmaLeftIdentity} is 
refined in the following graph:\\[0.25em]
\hspace*{0em}\begin{tabular}{rclcl}
$G_3$ & $=$ & $\{\ s\ \fleche{q}\ t\ |\ \exists\ a\ (s\ \fleche{a}_G\ t 
\,\wedge\ r\ \fleche{a}_G\ q)\ \}$ & $\cup$ & 
$\{\ r\ \fleche{s}\ s\ |\ s \in V_G - Q\ \}$\\[0.25em]
& $\cup$ & $\{\ s\ \fleche{r}\ s\ |\ s \in V_G-\{r\} \,\wedge \,r \not\in Q\ \}$
& $\cup$ & $\{\ s\ \fleche{t}\ r\ |\ s,t \in V_G - Q_r\ \}$\\[0.25em]
& $\cup$ & $\{\ q\ \fleche{s}\ t\ |\ \exists\ a\ 
(s\ \fleche{a}_G\ t \,\wedge\ r\ \fleche{a}_G\ q \neq r) \,\wedge 
\,s \not\in Q_r\ \}$
\end{tabular}\\[0.25em]
The last subset of \,$G_3$ \,is illustrated as follows:
\begin{center}
\includegraphics{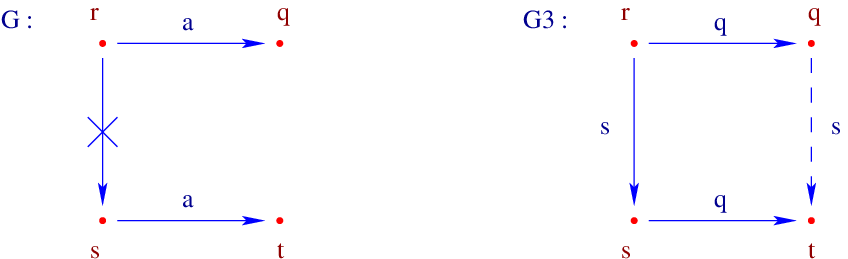}\\
\end{center}
Here is an example with \,$r \not\in Q$\,:
\begin{center}
\includegraphics{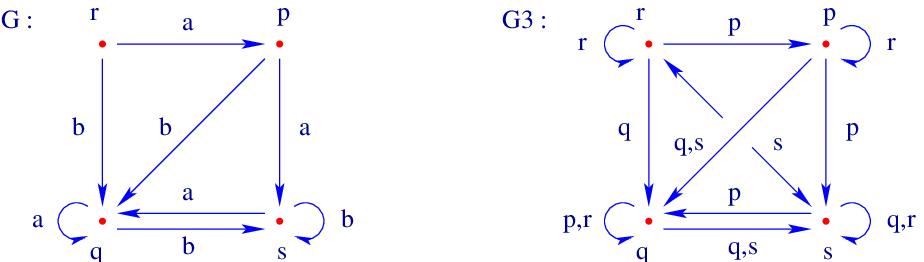}\\
\end{center}
The five subsets defining \,$G_3$ \,are disjoint. 
Thus \,$G_3$ \,remains deterministic.\\
Furthermore \,$G_3$ \,is source-complete and 
\,$V_{G_3} \,= \,V_G \,= \,A_{G_3}$\,.\\
Let \,$s \in V_G$\,. Let us check that \,$r\ \fleche{s}_{G_3}\ s$.\\
\hspace*{1em}Due to the second subset defining \,$G_3$\,, this is true when 
\,$s \not\in Q$.\\
\hspace*{1em}Assume that \,$s \in Q$. So \,$r\ \fleche{a}_G\ s$ \,for some 
\,$a \in A$.\\
\hspace*{1em}The first subset defining \,$G_3$ \,gives 
\,$r\ \fleche{s}_{G_3}\ s$.\\
So  \,$r$ \,is an out-simple $1$-root of \,$G_3$.\\
Let \,$s \in V_G$\,. Let us check that \,$s\ \fleche{r}_{G_3}\ s$.\\
\hspace*{1em}Due to the third subset defining \,$G_3$\,, this is true when 
\,$r \not\in Q$.\\
\hspace*{1em}Assume that \,$r \in Q$. So \,$r\ \fleche{a}_G\ r$ \,for some 
\,$a \in A$.\\
\hspace*{1em}As \,$r$ \,is loop-propagating, $s\ \fleche{a}_G\ s$. 
The first subset defining \,$G_3$ \,gives \,$s\ \fleche{r}_{G_3}\ s$.\\
By Lemma~\ref{LoopProp}, we just have to check that \,$r$ \,remains locally 
commutative for \,$G_3$ \,since by Lemma~\ref{EdgeProduct}, 
$(V_G,\croix_r)$ \,is a commutative magma of identity \,$r$ \,with 
\,$G_3 \,= \,{\cal C}(V_G)$ \,which implies that 
\,$G \,= \,{\cal C}\inter{V_G}^{|\interInd{Q}} \,= 
\,{\cal C}\inter{V_G\,,\,Q}$ \,with \,$\inter{q} = a$ \,for any 
\,$r\ \fleche{a}_G\ q$.\\
Let \,$r\ \fleche{xy}_{G_3}\ t$ \,for some \,$x,y \in V_G$\,. 
We have to show that \,$r\ \fleche{yx}_{G_3}\ t$.\\
As \,$r\ \fleche{x}_{G_3}\ x$ \,and \,$G_3$ \,is deterministic, we have 
\,$x\ \fleche{y}_{G_3}\ t$.\\
As \,$r\ \fleche{y}_{G_3}\ y$, we have to check that \,$y\ \fleche{x}_{G_3}\ t$.
\\
Let us start by checking it for \,$x = r$ \,or \,$y = r$.\\
Case $x = r$\,: \,So \,$r\ \fleche{y}_{G_3}\ t$ \,hence 
\,$y = t\ \fleche{r}_{G_3}\ t$.\\
Case $y = r$\,: \,So \,$x\ \fleche{r}_{G_3}\ t$ \,hence 
\,$y = r\ \fleche{x}_{G_3}\ x = t$.\\
It remains the four complementary cases below.\\
{\it Case 1}\,: $x,y \not\in Q_r$\,.\\
\hspace*{1em}So \,$x\ \fleche{y}_{G_3}\ t$ \,is only defined by the fourth 
subset hence \,$t = r$.\\
\hspace*{1em}By this fourth subset, we have \,$y\ \fleche{x}_{G_3}\ r = t$.\\
{\it Case 2}\,: $x \not\in Q_r$ \,and \,$y \in Q-\{r\}$.\\
\hspace*{1em}So \,$x\ \fleche{y}_{G_3}\ t$ \,is only defined by the first 
subset: we have \,$x\ \fleche{a}_G\ t$ \,for \,$r\ \fleche{a}_G\ y$.\\
\hspace*{1em}By the fifth subset, we get \,$y\ \fleche{x}_{G_3}\ t$.\\
{\it Case 3}\,: $x \in Q-\{r\}$ \,and \,$y \not\in Q_r$.\\
\hspace*{1em}So \,$x\ \fleche{y}_{G_3}\ t$ \,is only defined by the fifth 
subset: we have \,$y\ \fleche{a}_G\ t$ \,for \,$r\ \fleche{a}_G\ x$.\\
\hspace*{1em}By the first subset, we get \,$y\ \fleche{x}_{G_3}\ t$.\\
{\it Case 4}\,: $x,y \in Q-\{r\}$. 
There exists \,$a,b \in A$ \,such that \,$r\ \fleche{a}_G\ x$ \,and 
\,$r\ \fleche{b}_G\ y$.\\
\hspace*{1em}As \,$G$ \,is source-complete, there exists a vertex \,$z$ \,such 
that \,$x\ \fleche{b}_G\ z$.\\
\hspace*{1em}By the first subset, we have \,$x\ \fleche{y}_{G_3}\ z$. 
As \,$G_3$ \,is deterministic, $t = z$.\\
\hspace*{1em}So \,$r\ \fleche{ab}_G\ t$. As \,$r$ \,is locally commutative for 
\,$G$, we get \,$r\ \fleche{ba}_G\ t$.\\
\hspace*{1em}As \,$G$ \,is deterministic, $y\ \fleche{a}_G\ t$.
By the first subset, we have \,$y\ \fleche{x}_{G_3}\ t$.
\qed\\[1em]
For instance, let us consider the following deterministic and source-complete 
graph \,$G$\,:
\begin{center}
\includegraphics{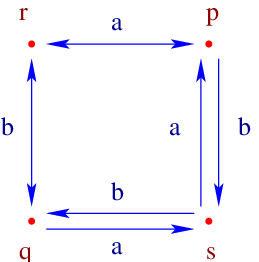}\\
\end{center}
which is also simple and loop-propagating. 
The vertex \,$r$ \,is locally commutative whereas \,$G$ \,is not a commutative 
graph since \,$p\ \fleche{ab}\ q$ \,and \,$p\ \fleche{ba}\ p$. 
By Theorem~\ref{MagmaCommutativity}, $G$ \,is a generalized Cayley graph of a 
commutative unital magma.\\
Precisely by Lemma~\ref{EdgeProduct}, 
$G \,= \,{\cal C}\inter{\{p,q,r,s\},\{p,q\}}$ \,where \,$\inter{p} \,= \,a$ 
\,and \,$\inter{q} \,= \,b$ \,for the commutative unital magma 
\,$(\{p,q,r,s\},\croix_r)$ \,defined by the following Cayley table:\\[0.5em]
\hspace*{16em}\begin{tabular}{|c||c|c|c|c|}
\hline
$\croix_r$ & $r$ & $p$ & $q$ & $s$ \\
\hline\hline
$r$ & $r$ & $p$ & $q$ & $s$ \\
\hline
$p$ & $p$ & $r$ & $s$ & $p$ \\
\hline
$q$ & $q$ & $s$ & $r$ & $q$ \\
\hline
$s$ & $s$ & $p$ & $q$ & $r$ \\
\hline
\end{tabular}

\section{Cayley graphs of monoids}\label{CayMon}

{\indent}We present graph-theoretic characterizations for the Cayley graphs 
of monoids, and of commutative monoids (see Theorem~\ref{Monoid}), and when 
they are left-cancellative and cancellative (see Theorems~\ref{MonoidBis} and 
\ref{MonoidCommutative}).\\[-0.5em]

Recall that a magma \,$(M,\cdot)$ \,is a {\it semigroup} \,if \,$\cdot$ \,is 
associative: \,$(p \cdot q) \cdot r \,= \,p \cdot (q \cdot r)$ \,for any 
\,$p,q,r \in M$. 
A semigroup with an identity element is a monoid. 
The identity is a propagating vertex for its generalized Cayley graphs.
\begin{fact}\label{OriginCayley}
For any generalized Cayley graph of a monoid, $1$ \,is propagating and 
out-simple.
\end{fact}
\proof\mbox{}\\
Let \,$G \,= \,{\cal C}\inter{M,Q}$ \,for some monoid \,$(M,\cdot)$ \,and 
\,$Q \subseteq M$.\\
By Fact~\ref{1Root}, the identity \,$1$ \,is an out-simple vertex. 
Let us check that \,$1$ \,is propagating.\\
Let \,$1\ \fleche{u,v}_G$ \,with \,$u,v \in A_G^*$\,.\\
So \,$u \,= \,\inter{p_1}\ldots\inter{p_m}$ \,and 
\,$v = \inter{q_1}\ldots\inter{q_n}$ \,for some \,$m,n \geq 0$ \,and 
\,$p_1,\ldots,p_m,q_1,\ldots,q_n \in Q$.\\
So \,$({\ldots}(1{\cdot}p_1){\ldots}){\cdot}p_m \,= 
\,({\ldots}(1{\cdot}q_1){\ldots}){\cdot}q_n$\,.
As \,$\cdot$ \,is associative,
\,$p_1{\cdot}{\ldots}{\cdot}p_m \,= \,q_1{\cdot}{\ldots}{\cdot}q_n$\,.\\
Let \,$s$ \,be any vertex of \,$G$. 
So \,$s\ \fleche{u}_G\ ({\ldots}(s{\cdot}p_1){\ldots}){\cdot}p_m \,= 
\,s{\cdot}(p_1{\cdot}{\ldots}{\cdot}p_m)$ \,and\\ 
\,$s\ \fleche{v}_G\ \,s{\cdot}(q_1{\cdot}{\ldots}{\cdot}q_n) \,= 
\,s{\cdot}(p_1{\cdot}{\ldots}{\cdot}p_m)$. 
Thus \,$s\ \fleche{u,v}_G$\,.
\qed\\[1em]
The commutativity of a semigroup is transposed on its generalized Cayley graphs.
\begin{fact}\label{CommutativeSemigroup}
Any generalized Cayley graph of a commutative semigroup is commutative.
\end{fact}
\proof\mbox{}\\
Let \,$G \,= \,{\cal C}\inter{M,Q}$ \,for some commutative semigroup \,$M$ 
\,and some \,$Q \subseteq M$.\\
Let us show that \,$G$ \,is commutative.\\
Let \,$s\ \fleche{\interFootnote{p}\interFootnote{q}}_G\ t$ \,for some 
\,$p,q \in Q$. By Lemma~\ref{Commutative}, it remains to check that 
\,$s\ \fleche{\interFootnote{q}\interFootnote{p}}_G\ t$.\\
We have \,$t \,= \,(s \cdot p) \cdot q \,= \,s \cdot (p \cdot q) \,= 
\,s \cdot (q \cdot p) \,= \,(s \cdot q) \cdot p$ \,hence 
\,$s\ \fleche{\interFootnote{q}\interFootnote{p}}_G\ t$.
\qed\\[1em]
When a monoid is left-cancellative, its generalized Cayley graphs are 
forward vertex-transitive.
\begin{fact}\label{LeftCancel}
Any generalized Cayley graph of a left-cancellative monoid is propagating.
\end{fact}
\proof\mbox{}\\
Let \,$G \,= \,{\cal C}\inter{M,Q}$ \,for some left-cancellative monoid 
\,$(M,\cdot)$ \,and some \,$Q \subseteq M$.\\
Let \,$s\ \fleche{u,v}_G$ \,with \,$u,v \in A_G^*$. Let \,$t \in M$. 
We have to check that \,$t\ \fleche{u,v}_G$\,.\\
So \,$u \,= \,\inter{p_1}\ldots\inter{p_m}$ \,and 
\,$v = \inter{q_1}\ldots\inter{q_n}$ \,for some \,$m,n \geq 0$ \,and 
\,$p_1,\ldots,p_m,q_1,\ldots,q_n \in Q$.\\
Thus \,$s{\cdot}(p_1{\cdot}{\ldots}{\cdot}p_m) \,= 
\,s{\cdot}(q_1{\cdot}{\ldots}{\cdot}q_n)$. 
As \,$\cdot$ \,is left-cancellative, 
$p_1{\cdot}{\ldots}{\cdot}p_m \,= \,q_1{\cdot}{\ldots}{\cdot}q_n$\,.\\
Hence \,$t\ \fleche{u,v}_G\ t{\cdot}(p_1{\cdot}{\ldots}{\cdot}p_m) \,= 
\,t{\cdot}(q_1{\cdot}{\ldots}{\cdot}q_n)$.
\qed\\[1em]
The {\it submonoid generated} \,by \,$Q \subseteq M$ \,is the least submonoid 
containing \,$Q$ \,{\it i.e.}\\
\hspace*{6em}$Q^* \,= \,\{\ q_1{\cdot}\ldots{\cdot}q_n \mid n \geq 0 \,\wedge 
\,q_1,\ldots,q_n \in Q\ \}$.
\begin{fact}\label{RootCayley}
A monoid \,$M$ \,is generated by \,$Q$ \ \ $\Longleftrightarrow$ \ \ 
$1$ \,is a root of \,${\cal C}\inter{M,Q}$.
\end{fact}
A {\it Cayley graph of a monoid} \,$M$ \,is a generalized Cayley graph 
\,${\cal C}\inter{M,Q}$ \,for \,$M$ \,generated by \,$Q$. 
The commutativity of such a graph coincides with the commutativity of \,$M$.
\begin{fact}\label{CayleyCommut}
A Cayley graph of a monoid \,$M$ \,is commutative \ ${\rm iff} \ M$ \,is 
commutative.
\end{fact}
\proof\mbox{}\\
$\Longleftarrow$\,: \,By Fact~\ref{CommutativeSemigroup}.\\
$\Longrightarrow$\,: \,Let a commutative graph 
\,$G \,= \,{\cal C}\inter{M,Q}$ \,for \,$M$ \,generated by \,$Q$. 
Let \,$s,t \in M$.\\
So \,$s \,= \,p_1{\cdot}{\ldots}{\cdot}p_m$ \,and 
\,$t \,= \,q_1{\cdot}{\ldots}{\cdot}q_n$ \,for some \,$m,n \geq 0$ \,and 
\,$p_1,\ldots,p_m,q_1,\ldots,q_n \in Q$.\\
So \,$1\ \fleche{u}_G\ s\ \fleche{v}_G\ s \cdot t$ \,for 
\,$u \,= \,\inter{p_1}{\cdot}{\ldots}{\cdot}\inter{p_m}$ \,and 
\,$v \,= \,\inter{q_1}{\cdot}{\ldots}{\cdot}\inter{q_n}$.\\
As \,$G$ \,is commutative, $1\ \fleche{vu}_G\ s \cdot t$.\\
As \,$G$ \,is deterministic and \,$1\ \fleche{v}_G\ t\ \fleche{u}_G\ t \cdot s$,
\,we get \,$s \cdot t \,= \,t \cdot s$.
\qed\\[1em]
In order to show that Facts~\ref{Magma}, \ref{OriginCayley}, \ref{RootCayley} 
characterize the Cayley graphs of monoids, we just have to apply 
Lemma~\ref{EdgeProduct} in the case where the root is propagating.
\begin{lemma}\label{PathProduct}
Let $r$ be a propagating root of a deterministic graph $G$.\\
\hspace*{1em}We can define the {\rm path-operation} \,$\ast_r$ \,for any 
\,$s,t \in V_G$ \,by\\
\hspace*{8em}$s\ \fleche{u}_G\ s \ast_r t$ \ if \ $r\ \fleche{u}_G\ t$ \,for 
some \,$u \in A^*$.\\
\hspace*{1em}Then \,$(V_G,\ast_r)$ \,is a monoid of identity \,$r$.\\
\hspace*{1em}If \,$r$ \,is commutative \,then \ $\ast_r$ \,is commutative.\\
\hspace*{1em}If \,$G$ \,is propagating \,then \ $\ast_r$ \,is 
left-cancellative.\\
\hspace*{1em}If \,$G$ \,is co-deterministic \,then \ $\ast_r$ \,is 
right-cancellative.\\
\hspace*{1em}If \,$G$ \,is target-complete \,then \,$\ast_r$ \,is 
left-invertible.\\
\hspace*{1em}If \,$r$ \,is a source-complete co-root \,then 
\,$\ast_r$ \,is right-invertible.\\
\hspace*{1em}If \,$r$ \,is source-complete \,then \,$\fleche{}_G(r)$ 
\,generates \,$V_G$ \,by \,$\ast_r$\,; moreover for \,$r$ \,out-simple,\\
\hspace*{8em}$G \,= \,{\cal C}\inter{V_G,\fleche{}_G(r)}$ \,where 
\,$\inter{q} \,= \,a$ \,for any \,$r\ \fleche{a}_G\ q$.
\end{lemma}
\proof\mbox{}\\
We have to apply Lemma~\ref{EdgeProduct} to the graph 
\,$\{\ s\ \fleche{\InfSupInd{u}}\ t\ |\ u \in A_G^* \,\wedge 
\,s\ \fleche{u}_G\ t\ \}$ \,where \,$\InfSup{\ } : A_G^*\ \fleche{}\ A$ \,is 
an injective mapping. The operation \,$\ast_r$ \,is illustrated as follows:
\begin{center}
\includegraphics{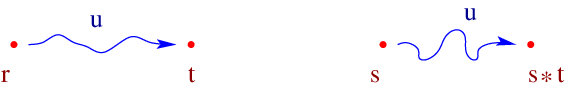}\\
\end{center}
Note that by Fact~\ref{SourcePropagating}, if \,$r$ \,is a source-complete 
vertex then \,$G$ \,is a source-complete graph.\\
Let us check the associativity of \,$\ast_r$ \,denoted by \,$\cdot$ \,in the 
sequel.\\
Let \,$x,y,z \in V_G$\,. 
\,We have to check that \,$(x{\cdot}y){\cdot}z \,= \,x{\cdot}(y{\cdot}z)$.\\
As \,$r$ \,is a root, there exists \,$v,w \in A_G^*$ \,such that 
\,$r\ \fleche{v}\ y$ \,and \,$r\ \fleche{w}\ z$.\\
Thus \,$x\ \fleche{v}\ x{\cdot}y\ \fleche{w}\ (x{\cdot}y){\cdot}z$ \ and \ 
$y\ \fleche{w}\ y{\cdot}z$. 
So \ $r\ \fleche{vw}\ y{\cdot}z$ \ \ hence \ \ 
$x\ \fleche{vw}\ x{\cdot}(y{\cdot}z)$.\\
As \,$G$ \,is deterministic, we get
\,$(x{\cdot}y){\cdot}z \,= \,x{\cdot}(y{\cdot}z)$.\\[0.25em]
It remains to check that \,$\fleche{}_G(r)$ \,is a generating subset of 
\,$V_G$ \,by \,$\cdot$\,. Let \,$s \in V_G$\,.\\
There exists a path 
\,$r = s_0\ \fleche{a_1}\ s_1{\ldots}s_{n-1}\ \fleche{a_n}\ s_n~=~s$.\\
As \,$r$ \,is source-complete, there exists \,$r_1,\ldots,r_n$ \,such 
that \,$r\ \fleche{a_1}\ r_1,\ldots,r\ \fleche{a_n}\ r_n$\,.\\
For every \,$1 \leq i \leq n$, $s_i \,= \,s_{i-1}{\cdot}r_i$ \,hence 
\,$s \,= \,(\ldots(r{\cdot}r_1)\cdot\ldots){\cdot}r_n \,= 
\,r_1\cdot\ldots{\cdot}r_n \,\in \,Q^*$.
\qed\\[1em]
We get a graph-theoretic characterization for the Cayley graphs of monoids.
\begin{theorem}\label{Monoid}
A graph is a Cayley graph of a $($resp. commutative$)$ monoid  \ 
{\rm if and only if} \ it is deterministic, source-complete with a propagating 
$($resp. and locally commutative$)$ out-simple root.
\end{theorem}
\proof\mbox{}\\
$\Longrightarrow$\,: \,By Facts~\ref{Magma}, \ref{OriginCayley}, 
\ref{RootCayley} (resp. and Fact~\ref{CommutativeSemigroup}).\\
$\Longleftarrow$\,: \,By Lemma~\ref{PathProduct} (resp. and 
Lemma~\ref{CommutativeForward}).
\qed\\[1em]
For instance, the following finite graph \,$G$\,:
\begin{center}
\includegraphics{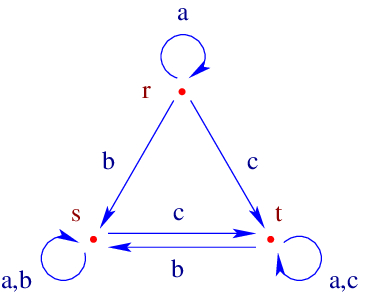}\\
\end{center}
is deterministic, source-complete, and \,$r$ \,is a propagating out-simple 
root. 
By Theorem~\ref{Monoid}, $G$ \,is a Cayley graph of a monoid. 
Precisely by Lemma~\ref{PathProduct}, 
$G \,= \,{\cal C}\inter{\{r,s,t\}}$ \,where \,$\inter{r} \,= \,a$\,, 
\,$\inter{s} \,= \,b$\,, \,$\inter{t} \,= \,c$ \,for the monoid 
\,$(\{r,s,t\},\ast_r)$ \,defined by the following Cayley table:\\[0.5em]
\hspace*{16em}\begin{tabular}{|c||c|c|c|}
\hline
$\ast_r$ & $r$ & $s$ & $t$ \\
\hline\hline
$r$ & $r$ & $s$ & $t$ \\
\hline
$s$ & $s$ & $s$ & $t$ \\
\hline
$t$ & $t$ & $s$ & $t$ \\
\hline
\end{tabular}\\[0.5em]
Another example is given by the following infinite graph:\\[0.25em]
\hspace*{1em}$G\ =\ \{\ n\ \fleche{a}\ n+2\ |\ n \in \entier\ \} \,\cup 
\,\{\ 2n\ \fleche{b}\ 2n+1\ |\ n \in \entier\ \} \,\cup 
\,\{\ 2n+1\ \fleche{a,b}\ 2n+3\ |\ n \in \entier\ \}$\\[0.25em]
represented as follows:
\begin{center}
\includegraphics{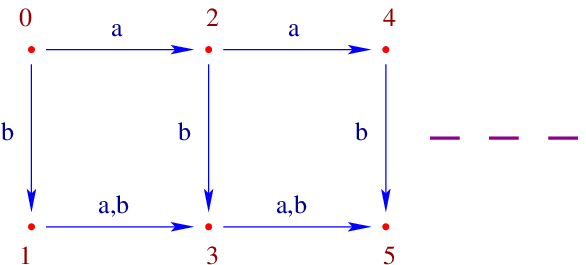}\\
\end{center}
Such a graph is deterministic, source-complete, and the root \,$0$ \,is 
out-simple, propagating and locally commutative. 
By Theorem~\ref{Monoid}, $G$ \,is a Cayley graph of a commutative monoid. 
Precisely by Lemma~\ref{PathProduct}, 
$G \,= \,{\cal C}\inter{\entier,\{1,2\}}$ \,where \,$\inter{1} \,= \,b$ \,and 
\,$\inter{2} \,= \,a$ \,for the path-operation \,$\ast$ \,from \,$0$ \,defined 
for any \,$p,q \in \entier$ \,by\\[0.25em]
\hspace*{10em}{$p \,\ast \,q \ = \ \left\{\begin{tabular}{ll}
$p+q$ & if \ $p$ \,or \,$q$ \,is even,\\[0.25em]
$p+q+1$ & if \ $p$ \,and \,$q$ \,are odd
\end{tabular}\right.$}\\[0.5em]
which is indeed a commutative monoid.\\
Let us adapt Theorem~\ref{Monoid} to right-cancellative monoids.
\begin{theorem}\label{RightMonoid}
A graph is a Cayley graph of a right-cancellative $($resp. commutative$)$ 
monoid \ {\rm if and only if} \ it is deterministic and co-deterministic, 
source-complete with a propagating $($resp. and locally commutative$)$ 
out-simple root.
\end{theorem}
\proof\mbox{}\\
$\Longrightarrow$\,: \,By Facts~\ref{Magma}, \ref{CancelMagma}, 
\ref{OriginCayley}, \ref{RootCayley} (resp. and 
Fact~\ref{CommutativeSemigroup}).\\
$\Longleftarrow$\,: \,By Lemma~\ref{PathProduct} (resp. and 
Lemma~\ref{CommutativeForward}).
\qed\\[1em]
By Theorem~\ref{RightMonoid}, the following graph:
\begin{center}
\includegraphics{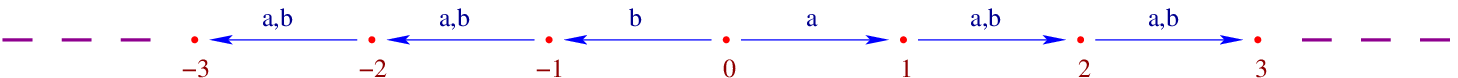}\\
\end{center}
is a Cayley graph of a right-cancellative monoid. 
By Lemma~\ref{PathProduct}, this graph is equal to 
\,${\cal C}\inter{\relatif,\{1,-1\}}$ \,where \,$\inter{1} \,= \,a$ \,and 
\,$\inter{-1} \,= \,b$ \,for the right-cancellative monoid \,$(\relatif,\ast)$ 
\,where the path-operation \,$\ast$ \,from \,$0$ \,is defined for any 
\,$m,n \in \relatif$ \,by\\[0.25em]
\hspace*{10em}{$m \,\ast \,n \ = \ \left\{\begin{tabular}{ll}
$m+|n|$ & if \ $m \geq 0$,\\[0.25em]
$m-|n|$ & if \ $m < 0$.
\end{tabular}\right.$}\\[0.5em]
We also deduce a known graph-theoretic characterization for the Cayley graphs 
of left-cancellative and cancellative monoids \cite{Cau}.
\begin{theorem}\label{MonoidBis}
A graph is a Cayley graph of a left-cancellative $($resp. cancellative$)$ 
monoid\\
{\rm iff} \ it is rooted, deterministic $($resp. and co-deterministic$)$, 
simple, propagating or forward vertex-transitive.
\end{theorem}
\proof\mbox{}\\
By Proposition~\ref{ForwardVertexTransitive}~(c) and for any deterministic
graph, it is equivalent for a graph to be propagating and forward vertex 
transitive.\\
$\Longrightarrow$\,: \,By Facts~\ref{Magma}, \ref{CancelMagma}, 
\ref{LeftCancel}, \ref{RootCayley}.\\
$\Longleftarrow$\,: \,By Fact~\ref{SourceCompletePropagating} and 
Lemma~\ref{PathProduct}.
\qed\\[1em]
Let us strengthen Theorem~\ref{MonoidBis} to the commutative case.
\begin{theorem}\label{MonoidCommutative}
For any graph \,$G$, the following properties are equivalent:\\
{\bf a)} \,$G$ \,is a Cayley graph of a left-cancellative 
$($resp. cancellative$)$ commutative monoid,\\
{\bf b)} \,$G$ \,is deterministic $($resp. and co-deterministic$)$, 
propagating with a locally commutative and out-simple root,\\
{\bf c)} \,$G$ \,is rooted, deterministic $($resp. and co-deterministic$)$, 
simple, forward vertex-transitive and commutative.
\end{theorem}
\proof\mbox{}\\
It suffices to apply Facts~\ref{SourceCompletePropagating} and 
\ref{CommutativeSemigroup} with Lemma~\ref{CommutativeForward} to 
Theorem~\ref{MonoidBis}.
\qed\\[1em]
For instance the following graph
\begin{center}
\includegraphics{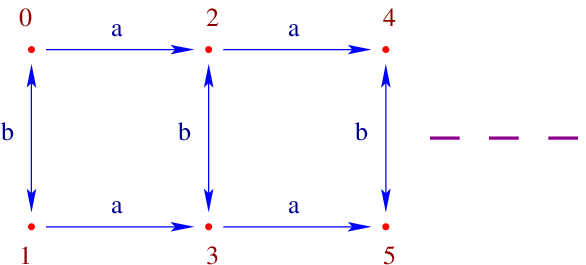}\\
\end{center}
is rooted, deterministic and co-deterministic, simple, forward 
vertex-transitive and commutative. 
By Theorem~\ref{MonoidCommutative}, it is a Cayley graph of a 
cancellative and commutative monoid. 
By Lemma~\ref{PathProduct}, this graph is equal to 
\,${\cal C}\inter{\entier,\{1,2\}}$ \,where \,$\inter{1} \,= \,b$ \,and 
\,$\inter{2} \,= \,a$ \,for the path-operation \,$\ast$ \,from \,$0$ \,defined 
for any \,$p,q \in \entier$ \,by\\[0.25em]
\hspace*{10em}{$p \,\ast \,q \ = \ \left\{\begin{tabular}{ll}
$p+q$ & if \ $p$ \,or \,$q$ \,is even,\\[0.25em]
$p+q-2$ & if \ $p$ \,and \,$q$ \,are odd
\end{tabular}\right.$}\\[0.5em]
which is indeed a cancellative and commutative monoid.

\section{Cayley graphs of semigroups and semilattices}

{\indent}We apply the previous characterizations for the Cayley graphs of 
monoids to the Cayley graphs of semigroups (see Theorem~\ref{Semigroup}) and 
of semilattices (see Theorem~\ref{Semilattice}).\\[-0.5em]

Recall that a {\it Cayley graph of a semigroup} \,$M$ \,is a generalized 
Cayley graph \,${\cal C}\inter{M,Q}$ \,such that \,$M \,= \,Q^+$ \,whose 
\,$Q^+ \,= \,\{\ q_1{\cdot}\ldots{\cdot}q_n \mid n > 0 \,\wedge 
\,q_1,\ldots,q_n \in Q\ \}$ \,is the {\it subsemigroup generated} \,by \,$Q$.
Let us extend Theorem~\ref{Monoid} into a characterization of these graphs.
\begin{theorem}\label{Semigroup}
A graph \,$G$ \,is a Cayley graph of a $($resp. commutative$)$ semigroup \ 
{\rm iff}\\
it is deterministic and there is an injection \,$i$ \,from \,$A_G$ \,into 
\,$V_G$ \,such that\\
\hspace*{2em}$G$ \,is accessible from \,$i(A_G)$,\\
\hspace*{2em}$i(a)\ \fleche{u}\ \inverse{v}\ i(b) \ \ \Longrightarrow \ \ 
s\ \fleche{au,bv}$ \ \ for any \,$s \in V_G$\,, \,$a,b \in A_G$ \,and 
\,$u,v \in A_G^*$\\
\hspace*{2em}$($resp. and \,$i(a)\ \fleche{b}\ \inverse{a}\ i(b)$ \,for any 
\,$a,b \in A_G)$.
\end{theorem}
\proof\mbox{}\\
$\Longrightarrow$\,: \,Let \,$G \,= \,{\cal C}\inter{M,Q}$ \,for some semigroup 
\,$(M,\cdot)$ \,generated by \,$Q$.\\
By Fact~\ref{Magma}, $G$ \,is deterministic.\\
If \,$M$ \,is not a monoid \,{\it i.e.} \,it has no identity, we turn \,$M$ 
\,into a monoid \,$M' \,= \,M \,\cup \,\{1\}$ \,by just adding an 
identity \,$1$ \,{\it i.e.} \,$p{\cdot}1 \,= \,1{\cdot}p \,= \,p$ \,for any 
\,$p \in M'$.\\
Let \,$M' \,= \,M$ \,when \,$M$ \,is a monoid.\\
In both cases, \,$M'$ \,is a monoid of identity \,$1$. 
Let \,$G' \,= \,{\cal C}\inter{M',Q} \,= 
\,G \,\cup \,\{\ 1\ \fleche{\interInd{q}}\ q\ |\ q \in Q\ \}$.\\
Let \,$i : A_G\ \fleche{}\ Q$ \,defined by \,$i(\inter{q}) = q$ \,for any 
\,$q \in Q$.\\
As \,$M$ \,is generated by \,$Q$, the graph \,$G$ \,is accessible from 
\,$Q \,= \,i(A_G)$.\\
Let \,$i(\inter{p})\ \fleche{u}_G\ _G\inverse{v}\ i(\inter{q})$ \,for some 
\,$p,q \in Q$ \,and \,$u,v \in A_G^*$\,.\\
So \,$p\ \fleche{u}_{G}\ _{G}\inverse{v}\ q$ \ hence \ 
$1\ \fleche{\interInd{p}u\,,\,\interInd{q}v}_{G'}$\,. 
By Fact~\ref{OriginCayley}, $1$ \,is a propagating vertex of \,$G'$.\\
Let \,$s \in V_G$\,. 
Therefore \,$s\ \fleche{\interInd{p}u\,,\,\interInd{q}v}_{G'}$ \ 
hence \ $s\ \fleche{\interInd{p}u\,,\,\interInd{q}v}_G$\,.\\
If \,$M$ \,is commutative, we have\\[0.25em]
\hspace*{6em}$i(\inter{p}) \,= \,p\ \fleche{\interInd{q}}_G\ p \cdot q \,= 
\,q \cdot p\ _G\inverse{\interInd{p}}\ q \,= \,i(\inter{q})$ \ for any
\,$p,q \in Q$.\\[0.25em]
$\Longleftarrow$\,: \,Let \,$G$ \,be a deterministic graph.\\
Let an injection \,$i : A_G\ \fleche{}\ V_G$ \,such that \,$G$ \,is accessible 
from \,$Q = i(A_G)$ \,and such that\\
\hspace*{1em}$s\ \fleche{au,bv}_G$ \,when 
\,$i(a)\ \fleche{u}_G\ _G\inverse{v}\ i(b)$ \,for any \,$s \in V_G$\,, 
\,$a,b \in A_G$ \,and \,$u,v \in A_G^*$\,.\hspace*{2em}(1)\\
Note that by (1), we get \,$s\ \fleche{a}_G$ \,for any \,$s \in V_G$ \,and 
\,$a \in A_G$\,, \,hence \,$G$ \,is source-complete.\\
We take a new vertex \,$r$ \,and we define the graph\\
\hspace*{9em}$\widehat{G} \,= 
\,G \,\cup \,\{\ r\ \fleche{a}\ i(a)\ |\ a \in A_G\ \}$.\\
So \,$\widehat{G}$ \,remains deterministic and source-complete, and \,$r$ \,is 
out-simple.\\
As \,$G$ \,is accessible from \,$Q$, \,$r$ \,is a root of \,$\widehat{G}$. 
By condition (1), $r$ \,is propagating.\\
By Lemma~\ref{PathProduct}, $V_{\widehat{G}} \,= \,V_G \cup \{r\}$ \,is a 
monoid for the path operation \,$\ast_r$ \,of identity \,$r$ \,and generated 
by \,$Q$ \,with \,$\widehat{G} \,= \,{\cal C}\inter{V_{\widehat{G}},Q} \,= 
\,{\cal C}\inter{V_{G},Q} \,\cup 
\,\{\ r\ \fleche{\interInd{q}}\ q\ |\ q \in Q\ \}$.\\
As \,$r$ \,is not the target of an edge of \,$\widehat{G}$ \,and by 
definition, $\ast_r$ \,remains an internal operation on \,$V_G$ \,{\it i.e.} 
\,$p \ast_r q \neq r$ \,for any \,$p,q \in V_G$\,. 
Thus \,$\ast_r$ \,remains associative on \,$V_G$\,.\\
Finally \,$G \,= \,\widehat{G}_{|V_G} \,= \,\,{\cal C}\inter{V_G\,,\,Q}$ \,and 
\,$(V_G\,,\,\ast_r)$ \,is a semigroup.\\
Let us assume that \,$i(a)\ \fleche{b}_G\ _G\inverse{a}\ i(b)$ \,for any 
\,$a,b \in A_G$\,.\\
So \,$r$ \,is locally commutative for \,$\widehat{G}$ \,and by 
Lemmas~\ref{CommutativeForward} and \ref{PathProduct}, $\ast_r$ \,is
commutative.
\qed\\[1em]
For instance, the two semilines \,$G_a \,= \,\{\ n\ \fleche{a}\ n+1\ |\ 
n \in \entier-\{0\}\ \} \,\cup \,\{\ n\ \fleche{a}\ n-1\ |\ 
n \in \relatif-\entier\ \}$ \,is a generalized Cayley graph of a monoid (see 
the previous example) but by Theorem~\ref{Semigroup}, $G_a$ \,is not a Cayley 
graph of a semigroup since it is not connected and has a unique label. 
On the other hand, $G_a \cup G_b$ \,is a Cayley graph of a (non commutative) 
semigroup. By Lemma~\ref{PathProduct}, 
$G_a \,\cup \,G_b \,= \,{\cal C}\inter{\relatif-\{0\},\{1,-1\}}$ \,where 
\,$\inter{1} \,= \,a$ \,and \,$\inter{-1} \,= \,b$ \,for the associative 
path-operation \,$\ast$ \,defined~by\\[0.25em]
\hspace*{8em}$m \ast n \,= \,m + sgn(m)\,|n|$ \ for any 
\,$m,n \in \relatif-\{0\}$\\[0.25em]
where \,$sgn(m) = 1$ \,for any $m > 0$ \,and \,$sgn(m) = -1$ \,for any $m < 0$.
\\[0.25em]
When a semigroup is cancellative, its generalized Cayley graphs are 
forward vertex-transitive.
\begin{fact}\label{Cancel}
Any generalized Cayley graph of a cancellative semigroup is propagating.
\end{fact}
\proof\mbox{}\\
Let \,$(M,\cdot)$ \,be a cancellative semigroup.\\
According to the proof of Fact~\ref{LeftCancel}, it remains to check that\\
\hspace*{10em}$s \,= \,s \cdot r \ \ \Longrightarrow \ \ t \,= \,t \cdot r$ 
\,for any \,$r,s,t \in M$.\\
Let \,$r,s,t \in M$ \,such that \,$s \,= \,s \cdot r$.
So \,$s \cdot s \,= \,s \cdot r \cdot s$.\\
As \,$\cdot$ \,is left-cancellative, \,$s \,= \,r \cdot s$. 
Thus \,$t \cdot s \,= \,t \cdot r \cdot s$.\\
As \,$\cdot$ \,is right-cancellative, we get \,$t \,= \,t \cdot r$.
\qed\\[1em]
Let us extend Theorem~\ref{MonoidBis} to the cancellative semigroups in order 
to get another description of their Cayley graphs \cite{Cau}.
\begin{theorem}\label{CancelSemigroup}
A graph \,$G$ \,is a Cayley graph of a cancellative $($resp. and commutative$)$ 
semigroup \ {\rm if and only if} \ 
it is deterministic and co-deterministic, and there is an injection~\,$i$ 
\,from \,$A_G$ \,into \,$V_G$ \,such that\\
\hspace*{2em}$G$ \,is accessible from \,$i(A_G)$,\\
\hspace*{2em}$i(a)\ \fleche{u}\ \inverse{v}\ i(b) \ \ \Longleftrightarrow \ \ 
i(c)\ \fleche{au,bv}$ \ \ for any \,$a,b,c \in A_G$ \,and \,$u,v \in A_G^*$\\
\hspace*{2em}$($resp. and \,$i(a)\ \fleche{b}\ \inverse{a}\ i(b)$ \,for any 
\,$a,b \in A_G)$.
\end{theorem}
\proof\mbox{}\\
Let us complete the proof of Theorem~\ref{Semigroup}.\\
$\Longrightarrow$\,: \,The monoid \,$M'$ \,remains cancellative (for instance 
see [Cau]).\\
By Fact~\ref{LeftCancel}, $G'$ \,is a propagating graph. 
Let \,$c \in A_G$\,. So \,$1 \,\downarrow_{G'} \,i(c)$.\\
Let \,$a,b \in A_G$ \,and \,$u,v \in A_G^*$\,. \,We have 
\,$1\ \fleche{au,bv}_{G'} \ \ \Longleftrightarrow \ \ 
i(c)\ \fleche{au,bv}_{G'}$\,.\\
Thus \,$i(a)\ \fleche{u}_G\ _G\inverse{v}\ i(b) \ \ \Longleftrightarrow \ \ 
i(c)\ \fleche{au,bv}_G$\,.\\[0.25em]
$\Longleftarrow$\,: \,Let us assume that\\
\hspace*{3em}$i(a)\ \fleche{u}_G\ _G\inverse{v}\ i(b) \ \ \Longleftrightarrow 
\ \ i(c)\ \fleche{au,bv}_G$ \ \ for any \,$a,b,c \in A_G$ \,and 
\,$u,v \in A_G^*$\,.\\
This equivalence coincides with\\
\hspace*{3em}$r\ \fleche{au,bv}_{\widehat{G}} \ \ \Longleftrightarrow \ \ 
i(c)\ \fleche{au,bv}_{\widehat{G}}$ \ \ for any \,$a,b,c \in A_G$ \,and 
\,$u,v \in A_G^*$\,.\\
As \,$r$ \,is target of no edge, we get\\
\hspace*{3em}$r\ \fleche{x,y}_{\widehat{G}} \ \ \Longleftrightarrow \ \ 
s\ \fleche{x,y}_{\widehat{G}}$ \ for any \,$r\ \fleche{}_{\widehat{G}}\ s$ \,and 
\,$x,y \in A_G^*$.\\
By Proposition~\ref{ForwardVertexTransitive}~(d), $\widehat{G}$ \,is a 
propagating graph.\\
By Lemma~\ref{PathProduct}, $\ast_r$ \,is left-cancellative on 
\,$V_{\widehat{G}} \,= \,V_G \,\cup \,\{r\}$ \,hence on \,$V_G$\,.\\
By hypothesis \,$G$ \,is co-deterministic. However \,$\widehat{G}$ \,can be not 
co-deterministic.\\
As \,$r$ \,is target of no edge, we get that \,$\ast_r$ \,is also 
right-cancellative.\\
\hspace*{1em}Precisely let \,$s,s',t \in V_G$ \,such that 
\,$s \ast_r t = s' \ast_r t$.\\
\hspace*{1em}There exists \,$u \in A_G^*$ \,such that 
\,$r\ \fleche{u}_{\widehat{G}}\ t$.\\ 
\hspace*{1em}So \ $s\ \fleche{u}_{\widehat{G}}\ s \ast_r t$ \ and \ 
$s'\ \fleche{u}_{\widehat{G}}\ s' \ast_r t \,= \,s  \ast_r t$.\\
\hspace*{1em}As \,$r$ \,is target of no edge, we have 
\,$s\ \fleche{u}_G\ s \ast_r t$ \ and \ $s'\ \fleche{u}_G\ s \ast_r t$.\\
\hspace*{1em}As \,$G$ \,is co-deterministic, we get \,$s = s'$.
\qed\\[1em]
For instance, let us consider the graph 
\,$G \,= \,\{\ (m,n)\ \fleche{p}\ (m+p,n+1)\ |\ m,n,p \in \entier\ \}$. 
By Theorem~\ref{CancelSemigroup}, this graph is a Cayley graph of a commutative 
and cancellative semigroup. By Lemma~\ref{PathProduct}, 
$G \,= \,{\cal C}\inter{\entier\croix\entier\,,\,\entier\croix\{0\}}$ \,where 
\,$\inter{(p,0)} \,= \,p$ \,for any \,$p \in \entier$, and for the 
path-operation \,$\ast$ \,defined by\\[0.25em]
\hspace*{8em}$(m,n) \ast (p,q) \,= \,(m+p,n+q+1)$ \ for any 
\,$m,n,p,q \in \entier$\\[0.25em]
which is indeed associative, cancellative and commutative.\\[0.25em]
A {\it semilattice} \,$(M,\cdot)$ \,is a commutative semigroup which is also 
{\it idempotent}\,: \,$p \cdot p = p$ \,for any \,$p \in M$. 
Let us apply Lemma~\ref{PathProduct} with Theorem~\ref{Semigroup}.
\begin{theorem}\label{Semilattice}
A graph \,$G$ \,is a Cayley graph of a semilattice \ {\rm if and only if}\\
it is deterministic and there is an injection \,$i$ \,from \,$A_G$ \,into 
\,$V_G$ \,such that\\
\hspace*{2em}$G$ \,is accessible from \,$i(A_G)$\\
\hspace*{2em}$i(a)\ \fleche{u}\ \inverse{v}\ i(b) \ \ \Longrightarrow \ \ 
s\ \fleche{au,bv}$ \ \ for any \,$s \in V_G$\,, $a,b \in A_G$ \,and 
\,$u,v \in A_G^*$\\
\hspace*{2em}$i(a)\ \fleche{b}\ \inverse{a}\ i(b)$ \ for any \,$a,b \in A_G$\\
\hspace*{2em}$i(a)\ \fleche{u}\ s \ \ \Longrightarrow \ \ s\ \fleche{au}\ s$ \ 
\ for any \,$s \in V_G$\,, \,$a \in A_G$ \,and \,$u \in A_G^*$.
\end{theorem}
\proof\mbox{}\\
Let us complete the proof of Theorem~\ref{Semigroup}.\\
$\Longrightarrow$\,: \,Suppose further that \,$s \in M$ \,is idempotent 
\,{\it i.e.} \,$s{\cdot}s \,= \,s$.\\
Let \,$i(\inter{q})\ \fleche{u}_G\ s$ \,with 
\,$u \,= \,\inter{q_1}{\cdot}{\ldots}{\cdot}\inter{q_n}$ \,for some 
\,$n \geq 0$ \,and \,$q,q_1,\ldots,q_n \in Q$.\\
As \,$i(\inter{q}) = q$, we get 
\,$s \,= \,q{\cdot}q_1{\cdot}\ldots{\cdot}q_n$ \,thus 
\,$s\ \fleche{\interInd{q}u}_G\ s{\cdot}q{\cdot}q_1{\cdot}\ldots{\cdot}q_n 
\,= \,s{\cdot}s \,= \,s$.\\[0.25em]
$\Longleftarrow$\,: \,Let \,$i(a)\ \fleche{u}\ s$ \,and \,$s\ \fleche{au}\ s$ 
\,for some \,$s \in V_G$\,, \,$a \in A_G$ \,and \,$u \in A_G^*$.\\
So \,$r\ \fleche{au}_{\widehat{G}}\ s$. 
Thus \,$s\ \fleche{au}_{\widehat{G}}\ s \,\ast_r \,s$.\\
As \,$r \not\in V_G$ \,is initial (not target of an edge), 
$s\ \fleche{au}_G\ s \ast_r s$.\\
As \,$G$ \,is deterministic, we get \,$s \ast_r s \,= \,s$.
\qed\\[1em]
For instance, the following finite graph:
\begin{center}
\includegraphics{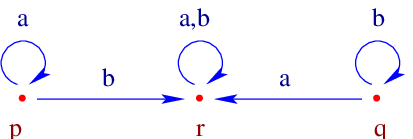}\\
\end{center}
satisfies the properties of Theorem~\ref{Semilattice} with \,$i(a) = p$ \,and
\,$i(b) = q$, hence is a Cayley graph of a semilattice. 
By Lemma~\ref{PathProduct}, this graph is equal to 
\,${\cal C}\inter{\{p,q,r\},\{p,q\}}$ \,where \,$\inter{p} \,= \,a$ \,and 
\,$\inter{q} \,= \,b$ \,for the semilattice \,$(\{p,q,r\},\ast)$ \,where 
\,$\ast$ \,is defined by the following Cayley table:\\[0.5em]
\hspace*{16em}\begin{tabular}{|c||c|c|c|}
\hline
$\ast$ & $p$ & $q$ & $r$ \\
\hline\hline
$p$ & $p$ & $r$ & $r$ \\
\hline
$q$ & $r$ & $q$ & $r$ \\
\hline
$r$ & $r$ & $r$ & $r$ \\
\hline
\end{tabular}

\section{Cayley graphs of groups}

{\indent}We have already given in \cite{Cau} a graph-theoretic 
characterization for the Cayley graphs of groups. 
We give here another description in a weak form. 
Finally, we strengthen these two characterizations for the Cayley graphs of 
abelian groups (see Theorem~\ref{GroupBis}).\\[-0.5em]

Recall that a {\it group} is a monoid left and right-invertible (w.r.t. the 
unit element). This induces the target completeness of its generalized Cayley 
graphs.
\begin{fact}\label{GroupTarget}
Any generalized Cayley graph of a group is target-complete.
\end{fact}
\proof\mbox{}\\
Let \,$G = {\cal C}\inter{M,Q}$ \,for some group \,$M$ \,and \,$Q \subseteq M$.
\\
We have \,$p {\cdot} q^{-1}\ \fleche{\interFootnote{q}}_G\ p$ \ for any 
\,$p \in M$ \,and \,$q \in Q$. Thus \,$G$ \,is target-complete.
\qed\\[1em]
Another basic property of the generalized Cayley graphs follows from 
Fact~\ref{OriginCayley}.
\begin{fact}\label{Group}
For any generalized Cayley graph of a group, $1$ \,is a chain-propagating 
vertex.
\end{fact}
\proof\mbox{}\\
Let \,$G = {\cal C}\inter{M,Q}$ \,for some group \,$M$ \,and \,$Q \subseteq M$. 
For any \,$q \in Q$, we have\\[0.25em]
\hspace*{6em}$s\ \fleche{\overline{\interInd{q}}}_G\ t \ \ \Longleftrightarrow 
\ \ t\ \fleche{\interInd{q}}_G\ s \ \ \Longleftrightarrow \ \ s = t \cdot q \ \ 
\Longleftrightarrow \ \ t = s \cdot q^{-1}$.\\[0.25em]
Let \,$H \,= \,{\cal C}\inter{M,Q \cup Q^{-1}}$ \,for \,$\inter{\ }$ 
\,extended injectively on \,$Q^{-1}-Q$. We have\\
\hspace*{6em}$s\ \fleche{u}_G\ t \ \ \Longleftrightarrow \ \ 
s\ \fleche{i(u)}_H\ t$ \ for any \,$u \in (Q \cup \overline{Q})^*$ \,and 
\,$s,t \in M$\\
where \,$i$ \,is the morphism on \,$(Q \cup \overline{Q})^*$ \,into 
\,$(Q \cup Q^{-1})^*$ \,defined by\\
\hspace*{10em}$i(q) = q$ \ and \ $i(\overline{q}) = q^{-1}$ \ for any 
\,$q \in Q$.\\
By Fact~\ref{OriginCayley}, $1$ \,is a propagating vertex 
of \,$H$ \,{\it i.e.} \,$1$ \,is a chain-propagating vertex of \,$G$.
\qed\\[1em]
Recall that a {\it Cayley graph of a group} \,$M$ \,is a generalized Cayley 
graph \,${\cal C}\inter{M,Q}$ \,such that \,$M$ \,is generated by \,$Q$ 
\,{\it i.e.} \,$M$ \,is equal to the least subgroup 
\,$(Q \,\cup \,Q^{-1})^*$ \,containing \,$Q$ \,where 
\,$Q^{-1} \,= \,\{\ q^{-1}\ |\ q \in Q\ \}$ \,is the set of inverses of the 
elements in \,$Q$.
\begin{fact}\label{ConnectedCayley}
A group \,$M$ \,is generated by \,$Q$ \ \ $\Longleftrightarrow$ \ \ 
${\cal C}\inter{M,Q}$ \,is connected.
\end{fact}
Let us adapt Theorem~\ref{MonoidBis} for groups. 
We extend Lemma~\ref{PathProduct}.
\begin{proposition}\label{ChainProduct}
Let \,$r$ \,be a chain-propagating vertex of a connected, deterministic and\\
\hspace*{1em}co-deterministic graph \,$G$. 
We can define the {\rm chain-operation} \,$\overline{\ast}_r$ 
\,for any \,$s,t \in V_G$ \,by\\
\hspace*{8em}$s\ \fleche{u}_G\ s \,\overline{\ast}_r \,t$ \ if \ 
$r\ \fleche{u}_G\ t$ \,for some \,$u \in (A_G \cup \overline{A_G})^*$.\\
\hspace*{1em}Then \,$(V_G,\overline{\ast}_r)$ \,is a right-cancellative monoid 
of identity \,$r$.\\
\hspace*{1em}If \,$r$ \,is chain-commutative then \,$\overline{\ast}_r$ \,is 
commutative.\\
\hspace*{1em}If \,$r$ \,is source and target-complete then 
\,$(V_G,\overline{\ast}_r)$ \,is a group generated by \,$\fleche{}_G(r)$\,;\\
\hspace*{2em}moreover if \,$r$ \,is an in-simple and out-simple vertex \,then\\
\hspace*{8em}$G \,= \,{\cal C}\inter{V_G,\fleche{}_G(r)}$ \,where 
\,$\inter{q} \,= \,a$ \,for any \,$r\ \fleche{a}_G\ q$.
\end{proposition}
\proof\mbox{}\\
{\bf i)} The graph \,$\overline{G}$ \,is strongly connected and it remains 
deterministic and co-deterministic.\\
Furthermore \,$r$ \,is a propagating vertex of \,$\overline{G}$.\\
By Lemma~\ref{PathProduct}, $(V_{\overline{G}}\,,\ast_r)$ \,is a 
right-cancellative monoid of identity \,$r$.\\
For any \,$u \in (A_G \cup \overline{A_G})^*$, the binary relation 
\,$\fleche{u}_G$ \,on \,$V_G = V_{\overline{G}}$ \,is equal to 
\,$\fleche{u}_{\overline{G}}$\,.\\
Thus \,$_{\overline{G}}\ast_r$ \,is equal to \,$_G\overline{\ast}_r$\,. 
This operation \,$\overline{\ast}_r$ \,is illustrated as follows:
\begin{center}
\includegraphics{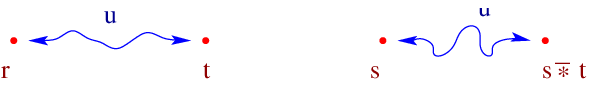}\\
\end{center}
If \,$r$ \,is chain-commutative for \,$G$ \,then \,$r$ \,is commutative for 
\,$\overline{G}$,\\
hence by Lemma~\ref{PathProduct}, 
$_{\overline{G}}\ast_r$ \,is commutative.\\[0.25em]
{\bf ii)} Let us assume that \,$r$ \,is source-complete and target-complete.\\
By Fact~\ref{SourcePropagating}, $G$ \,is source and target-complete.\\
So \,$\overline{G}$ \,remains source and target-complete.\\
By Lemma~\ref{PathProduct}, $(V_G\,,\ast_r)$ \,is a group generated by
\,$\fleche{}_G(r)$.\\
Suppose that in addition \,$r$ \,is an in-simple and out-simple vertex of 
\,$G$. 
However \,$r$ \,can be not an out-simple vertex of \,$\overline{G}$.
As in the proof of Fact~\ref{Group}, we restrict \,$\overline{G}$ \,to the 
graph\\[0.25em]
\hspace*{10em}$\widehat{G} \ = \ G \,\cup \,\{\ t\ \fleche{\overline{a}}\ s\ |\ 
s\ \fleche{a}_G\ t\ \nofleche{}{}\,\!_G\ s\ \}$.\\[0.25em]
So \,$\widehat{G}$ \,is the label restriction of \,$\overline{G}$ \,to 
\,${A_{\widehat{G}}}$\,.\\
Thus \,$\widehat{G}$ \,remains deterministic and \,$r$ \,is a source-complete 
propagating root of \,$\widehat{G}$.\\
Furthermore \,$r$ \,is an out-simple vertex of \,$\widehat{G}$.\\
By Lemma~\ref{PathProduct}, 
\,$\widehat{G} \,= \,{\cal C}\inter{V_G,\fleche{}_{\widehat{G}}(r)}$ \ with \ 
$\inter{s} \,= \,a$ \ for any \,$r\ \fleche{a}_{\widehat{G}}\ s$.\\
Precisely for any \,$s \in \fleche{}_{\widehat{G}}(r) \,= 
\,\fleche{}_G(r) \cup \fleche{}_{G^{-1}}(r)$, we have\\[0.25em]
\hspace*{10em}{$\inter{s} \ = \ \left\{\begin{tabular}{ll}
$a$ & if \ $r\ \fleche{a}_G\ s$\\[0.25em]
$\overline{a}$ & if \ $s\ \fleche{a}_G\ r$.
\end{tabular}\right.$}\\[0.25em]
Finally \,$G \,= \,\widehat{G}^{|A_G} \,= \,{\cal C}\inter{V_G,\fleche{}_G(r)}$.
\qed\mbox{}\\[1em]
We can complete a graph-theoretic characterization for the Cayley graphs of 
groups \cite{Cau}.
\begin{theorem}\label{GroupBis}
For any graph \,$G$, the following properties are equivalent:\\
{\bf a)} \,$G$ \,is a Cayley graph of a $($resp. commutative$)$ group,\\
{\bf b)} \,$G$ \,is connected, deterministic and co-deterministic, 
with a chain-propagating $($resp. and locally commutative$)$ source and 
target-complete in-simple and out-simple vertex,\\
{\bf c)} \,$G$ \,is connected, simple, deterministic and co-deterministic, 
vertex-transitive $($resp. and commutative$)$.
\end{theorem}
\proof\mbox{}\\
$(a)\ \Longrightarrow\ (b)$\,: \,Let \,$G$ \,be a Cayley graph of a group.\\
By Fact~\ref{Magma}, \,$G$ \,is deterministic and source-complete.\\
By Fact~\ref{CancelMagma}, \,$G$ \,is co-deterministic and simple.\\
By Fact~\ref{GroupTarget}, $G$ \,is target-complete.
By Fact~\ref{Group}, the identity \,$1$ \,is a chain-propagating vertex.\\
By Fact~\ref{ConnectedCayley}, $G$ \,is connected.\\
If the group is commutative then by Fact~\ref{CommutativeMagma}, the identity 
\,$1$ \,is also locally commutative.\\
$(b)\ \Longrightarrow\ (a)$\,: \,By Fact~\ref{SourcePropagating} and 
Proposition~\ref{ChainProduct} 
(resp. and Lemma~\ref{PropagatingChainCommutative}).\\[0.5em]
$(b)\ \Longleftrightarrow\ (c)$\,: \,By Fact~\ref{SourcePropagating} and 
Corollary~\ref{VertexTransitive}~(e) (resp. and 
Lemma~\ref{PropagatingChainCommutative}).
\qed\mbox{}\\[1em]
Let us generalize the graph defined after Theorem~\ref{MonoidCommutative} into 
the following graph:
\begin{center}
\includegraphics{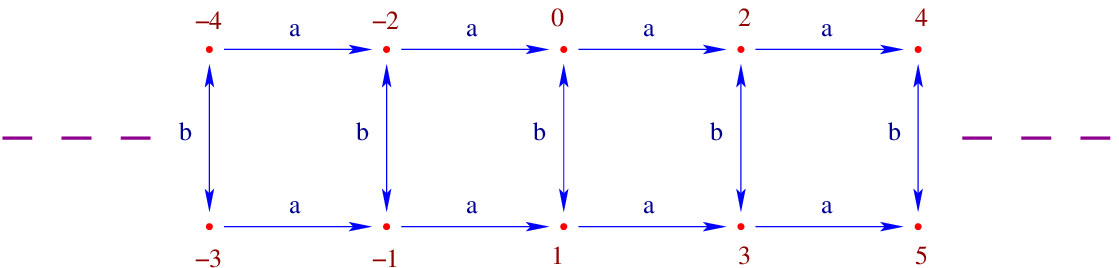}\\
\end{center}
This graph satisfies the properties of Theorem~\ref{GroupBis}~(b) or (c). 
Thus and according to Proposition~\ref{ChainProduct}, the resulting 
chain-operation \,$\overline{\ast}$ \,defined for any \,$p,q \in \relatif$ \,by
\\[0.25em]
\hspace*{10em}{$p \,\overline{\ast} \,q \ = \ \left\{\begin{tabular}{ll}
$p+q$ & if \ $p$ \,or \,$q$ \,is even,\\[0.25em]
$p+q-2$ & if \ $p$ \,and \,$q$ \,are odd
\end{tabular}\right.$}\\[0.5em]
makes that \,$(\relatif,\overline{\ast})$ \,is a commutative (abelian) group. 
Similarly, the following graph:
\begin{center}
\includegraphics{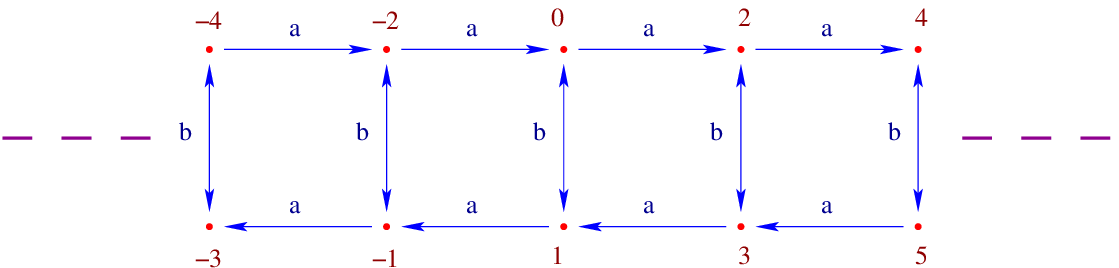}\\
\end{center}
is a Cayley graph of a (non commutative) group.

\section{Conclusion}

We obtained simple graph-theoretic characterizations for Cayley graphs of 
elementary algebraic structures. 
This is a first step in a graph description of algebraic structures.

\bibliographystyle{alpha}

\end{document}